\documentclass[preprint,12pt]{elsarticle}
\usepackage{graphicx}
\usepackage{epstopdf, epsfig}

\usepackage{natbib}
\usepackage{hyperref}
\usepackage{amsmath,amssymb,subcaption,float}
\usepackage{booktabs}
\usepackage{multirow}
\usepackage{comment}
\usepackage{xcolor}
\newcommand{\rev}[1]{\textcolor{red}{#1}}
\usepackage{adjustbox}
\usepackage{tabularx}			
\usepackage{longtable}
\usepackage{nicematrix}
\usepackage{array}
\usepackage{makecell}
\usepackage{tabularray}
\usepackage{physics}
\usepackage{bm}
\usepackage{mathrsfs}
\usepackage{svg}
\journal{Journal of Computational Physics}

\begin{document}
\begin{frontmatter}

\title{Body-Free Simulation of Three-Dimensional Turbulent Cylinder Wakes}

\author[aff1]{Zhicheng Wang}
\author[aff2]{Theo K\"aufer}
\author[aff1]{Khemraj Shukla}
\author[aff2]{Michael Triantafyllou}
\author[aff1]{George Em Karniadakis\corref{cor1}}
\cortext[cor1]{Corresponding author.}
\ead{George_Karniadakis@brown.edu}

\affiliation[aff1]{organization={Division of Applied Mathematics, Brown University},
            city={Providence}, state={RI}, postcode={02906}, country={USA}}
\affiliation[aff2]{organization={Department of Mechanical Engineering, Massachusetts Institute of Technology},
            city={Cambridge}, state={MA}, postcode={02139}, country={USA}}

\begin{abstract}
\rev{We present a body-free simulation framework for three-dimensional turbulent cylinder wakes, in which the upstream cylinder is not explicitly resolved. Instead, the incompressible Navier--Stokes equations are solved in a simplified rectangular domain, and the inflow is prescribed using velocity profiles extracted from experimental measurements or pre-computed direct numerical simulations (DNS). We show that, for the Reynolds numbers considered here, prescribing low-dimensional inflow information at a single downstream location is sufficient to reconstruct the principal wake dynamics, including three-dimensionality, coherent vortex shedding, Reynolds-stress distributions, and spectral content, for $Re=500$, $5{,}000$, and $11{,}000$. Comparisons with full-body DNS and particle image velocimetry (PIV) measurements show good agreement in mean velocity profiles, Reynolds stresses, and dominant shedding frequencies. A local stability analysis further shows that successful wake reconstruction is obtained when the imposed inflow is selected from the absolutely unstable near-wake region. The results support existing theory that the essential dynamics of bluff-body wakes are governed primarily by the instability of the near-wake profile rather than by the explicit presence of the body itself. The role of the onset of three-dimensionality is elucidated for first time. Relative to full DNS with the cylinder present, the proposed body-free simulation offers a substantial reduction in computational cost, providing a physically interpretable and computationally efficient route for wake reconstruction, reduced-complexity simulation, and flow-control studies.}
\end{abstract}


\end{frontmatter}

\section{Introduction}
The unsteady wake behind bluff bodies such as circular cylinders is a canonical problem in fluid mechanics, revealing fundamental mechanisms of vortex shedding, instability, and transition. Despite its apparent geometric simplicity, the turbulent wake exhibits complex three-dimensional dynamics that depend sensitively on Reynolds number, body shape, and inflow conditions \cite{Williamson1996Vortex,Roshko1954Cylinder,MOIN_cylinder_PoF2000}. Conventional direct numerical simulations (DNS) must include the body itself in order to generate the wake, which substantially increases the computational cost and makes it more difficult to isolate the intrinsic instability mechanisms of the wake.

Pioneering linear-stability analyses established that the formation of the von K\'arm\'an vortex street is governed primarily by the absolute instability of the mean velocity profile in the near wake \cite{Triantafyllou1986Formation}. Building on this understanding, \cite{KARNIDAKIS_reduced_model_1991} introduced a reduced-order model in which the Navier--Stokes equations were solved without the upstream cylinder, using as inflow a time-averaged velocity profile extracted from a pre-computed DNS at $Re=100$. Remarkably, that minimal model reproduced the two-dimensional vortex street and its shedding frequency, demonstrating that the essential wake dynamics are self-sustaining once a sufficiently unstable mean profile is prescribed at the inlet. \cite{Chan_reduced_model_2006} likewise reproduced the dynamics of shallow vortex streets by prescribing only a time-averaged velocity profile at a downstream location. Separately, the secondary-instability analysis of \cite{Henderson1996_PoF} showed that the cylinder wake becomes three-dimensional above $Re=188.5$. However, to the best of our knowledge, no previous study has examined whether a body-free simulation can also reproduce this transition to a fully three-dimensional wake.

\rev{In the present work, we extend this body-free simulation concept to the three-dimensional turbulent regime and demonstrate that a properly defined inflow condition---obtained either from experimental particle image velocimetry (PIV) data or from high-fidelity DNS---can accurately generate the full 3D wake in the absence of the upstream body. Specifically, for the Reynolds numbers considered here, prescribing low-dimensional inflow information at a single downstream location is sufficient to reproduce the principal wake structure, including coherent vortical motions, Reynolds stresses, and spectral energy distributions. The resulting flow fields closely match those of the corresponding full-body DNS and experiments, supporting the interpretation that the dominant wake dynamics are governed primarily by the instability of the near-wake profile rather than by the explicit presence of the body.}

\rev{More broadly, the present work is related to body-exclusive and downstream-initialized wake simulations, in which one studies wake evolution without explicitly resolving the wake-generating body. Such strategies have appeared in far-wake DNS/LES and other bluff-body wake contexts, where the downstream flow is initialized from precursor simulations or prescribed velocity fields. The present study addresses a different computational question: whether a spatially developing three-dimensional turbulent cylinder wake can be regenerated in an inflow--outflow simulation from only low-dimensional near-wake information, and whether the success of that regeneration can be predicted from the instability properties of the imposed profile. In this sense, the contribution is not merely body exclusion, but the combination of body-free wake reconstruction, physics-based inflow selection through local absolute-instability analysis, and systematic assessment of the role of the crossflow component using DNS- and PIV-derived inflow data.}

\rev{The main contributions of this work are threefold. First, we demonstrate that a body-free simulation framework can reconstruct the principal dynamics of a fully three-dimensional turbulent cylinder wake, extending earlier two-dimensional formulations to substantially more complex flow regimes. Second, we show that this reconstruction can be driven not only by DNS-derived inflow data but also by experimentally measured PIV profiles, thereby linking the method directly to laboratory data. Third, we show that local absolute-instability analysis provides a practical physics-based criterion for selecting inflow locations, and we further use controlled modifications of the inflow crossflow component to identify how transverse momentum input affects frequency selection and wake three-dimensionality.}

This body-free simulation framework provides a reduced-complexity approach for reconstructing and studying turbulent wakes directly from limited inflow data. It also opens new opportunities for data-driven model reduction, flow control, and the physical interpretation of self-sustained instabilities in separated flows.
The remainder of the paper is organized as follows. Section~\ref{sec:results_stability} presents the local linear stability analysis used to identify the near-wake region that remains absolutely unstable. Section~\ref{sec:reduced_model} describes the body-free simulation setup and the generation of the inflow data. The results of the body-free simulations and their comparison with DNS and PIV measurements are presented in Section~\ref{sec:results_statistics} and Section~\ref{sec:results_cross-flow}. Finally, conclusions are given in Section~\ref{sec:conclusion}.

\section{\rev{Local linear stability analysis of the mean wake}}
\label{sec:results_stability}
\rev{To characterize the instability properties of the cylinder wake, we perform a local linear stability analysis based on the time-averaged mean flow obtained from DNS. Following \cite{Huerre_AnnRev_Stability}, local instability analysis in a spatially developing wake is justified when the base flow varies slowly in the streamwise direction, allowing each streamwise station to be treated as a locally parallel profile. The mean streamwise velocity profile $\overline{U}(y)$ is extracted over the streamwise range $x/D\in[0.6,2.0]$ for flow past a stationary circular cylinder at $Re=500$, $5\times10^3$, and $1.1\times10^4$. These profiles are used as base states for a local parallel-flow stability analysis. Specifically, we consider the inviscid Orr-Sommerfeld equation or Rayleigh equation \cite{Triantafyllou1986Formation}}
\begin{equation}
    (k \overline{U}-\omega)(\ddot{\Psi}-k^2\Psi)=k\ddot{\overline{U}}\Psi,
\end{equation}
\rev{where $k$ and $\omega$ are the complex wavenumber and frequency, respectively, and $\Psi$ is the streamfunction. Here, $\ddot{\Psi}$ denotes the second derivative with respect to $y$. Within this framework, a profile is classified as absolutely unstable when the critical-point frequency satisfies $\Im(\omega)>0$, and convectively unstable when $\Im(\omega)<0$. The purpose of this analysis is not to provide a quantitatively exact prediction of the fully three-dimensional turbulent wake, but rather to identify the streamwise region in which the mean profile retains sufficient intrinsic instability to sustain vortex formation in the absence of the bluff body. For each $x/D$ location, the corresponding mean profile is extracted over $y/D\in[-5,5]$.}

\rev{In the present work, the Rayleigh equation is discretized using a five-point centered finite-difference scheme. The stability problem is solved on the half-domain $y/D\ge 0$, taking advantage of the symmetry of the time-averaged wake with respect to the centerline. In the interior of the computational domain, the second derivative operator is approximated by a fourth-order five-point stencil, while near-boundary rows are treated with lower-order centered approximations to maintain a consistent matrix form.}

\rev{To construct the spatio-temporal stability maps, a sequence of complex wavenumbers is considered by fixing several values of $k_i$ and varying $k_r$ over the interval $k_r \in [0.2,4.0]$. For each complex wavenumber $k=k_r+ik_i$, the generalized eigenvalue problem is solved and the resulting eigenvalues are mapped in the complex $\omega$-plane. The physically relevant branch is then followed continuously by root-tracking between neighboring wavenumbers using eigenvector overlap, which allows the same mode to be identified as $k$ is varied. Repeating this procedure for multiple values of $k_i$ yields families of contours in the $\omega$-plane, from which the pinch point associated with the absolute mode can be identified.}

\begin{figure}
\centering
  \begin{subfigure}[t]{0.425\textwidth}
    \includegraphics[width=\textwidth,trim=10 10 10 10,clip]{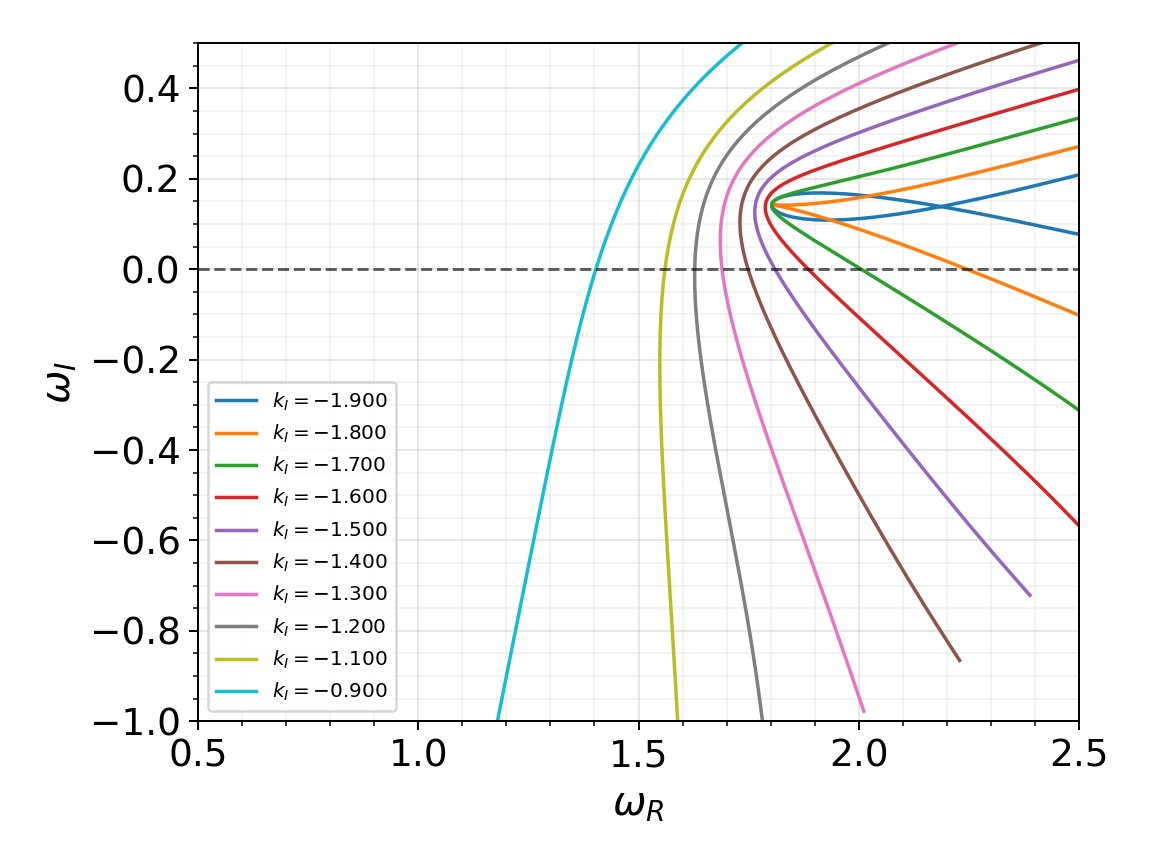}
        \label{fig:st_re5k_x08}
      \vspace*{-5mm}
    \subcaption*{$x/D=0.8$}
  \end{subfigure}
  \hspace{5mm}
  \begin{subfigure}[t]{0.425\textwidth}
    \includegraphics[width=\textwidth,trim=10 10 10 10,clip]{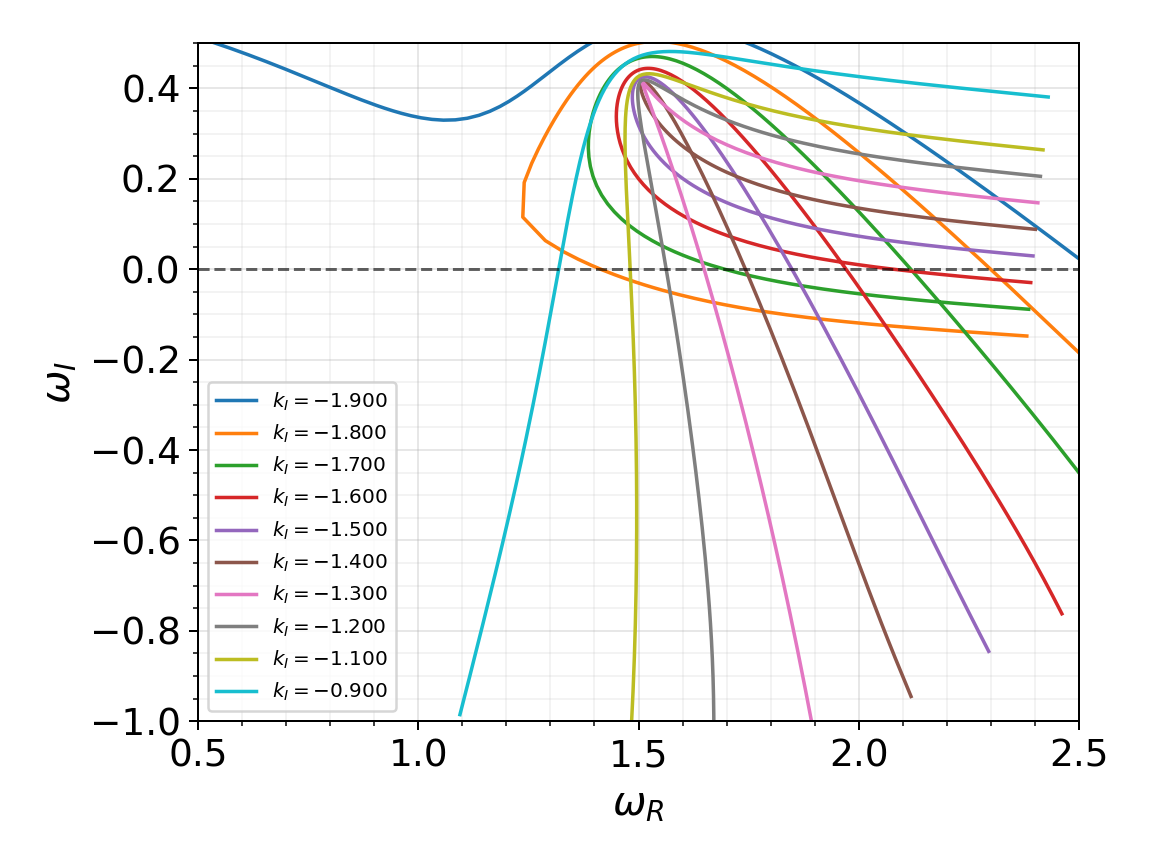}
    \label{fig:st_re5k_x12}
      \vspace*{-5mm}
     \subcaption*{$x/D=1.2$}
      \end{subfigure}   
         \begin{subfigure}[t]{0.425\textwidth}
    \includegraphics[width=\textwidth,trim=10 10 10 0,clip]{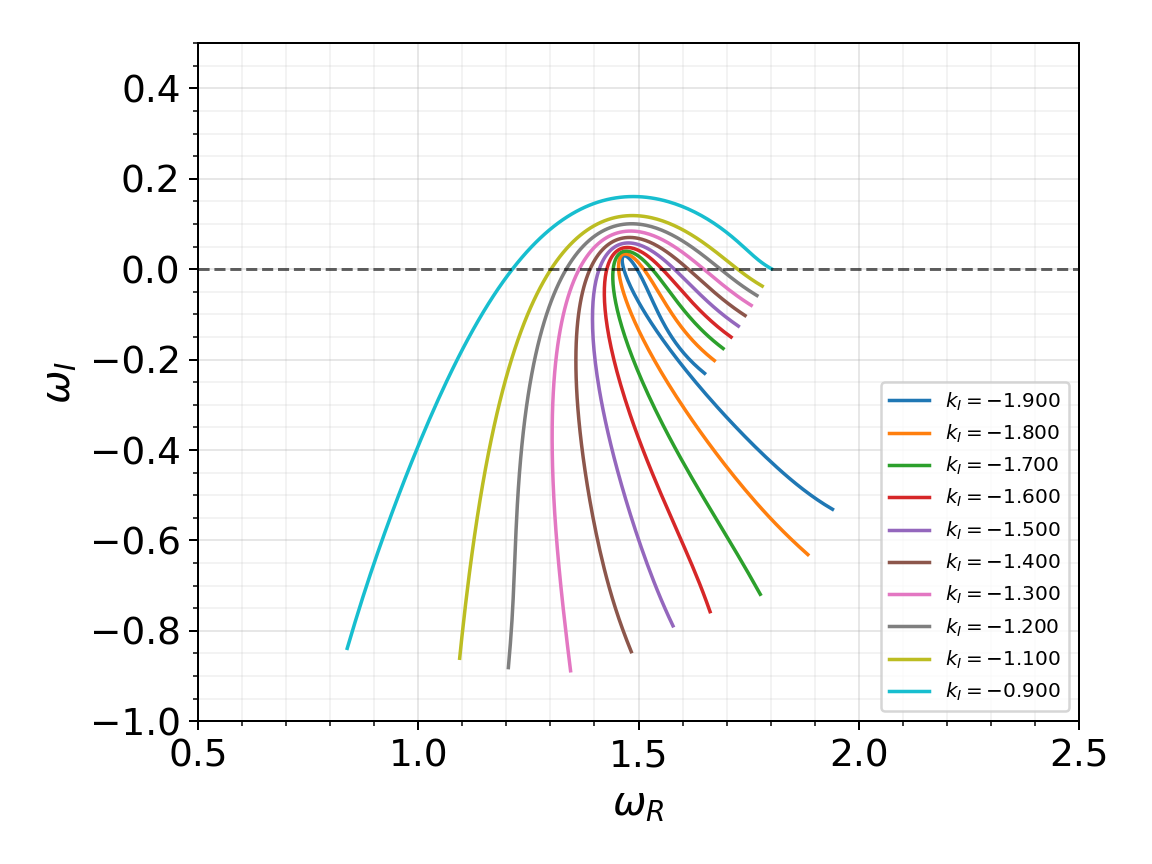}
        \label{fig:st_re5k_x16}
      \vspace*{-5mm}
    \subcaption*{$x/D=1.6$}
  \end{subfigure}
  \hspace{5mm}
        \begin{subfigure}[t]{0.425\textwidth}
    \includegraphics[width=\textwidth,trim=10 10 10 0,clip]{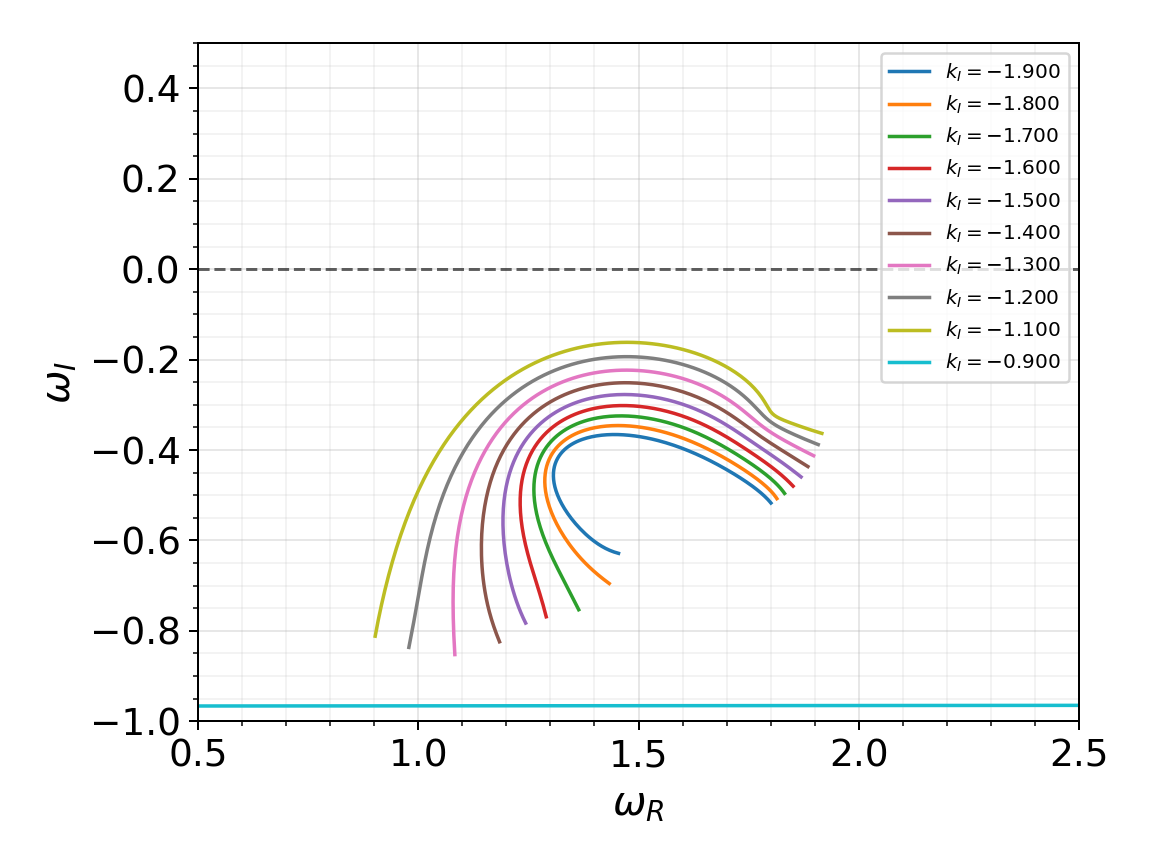}
    \label{fig:st_re5k_x18}
      \vspace*{-5mm}
     \subcaption*{$x/D=1.8$}
      \end{subfigure}    
   \caption{Spatio-temporal stability contours in the $\omega$-plane for the mean wake at $Re=5\times 10^3$ and several streamwise locations. Each contour corresponds to a fixed imaginary part of the wavenumber, $k_i$, while $k_r$ is varied in $[0.2,4.0]$. The critical point structure identifies the absolute mode of the local mean profile; its progressive weakening downstream indicates the gradual loss of absolute instability.}
       \label{fig:omega_plane_re5k}
\end{figure}

\rev{Figure~\ref{fig:omega_plane_re5k} illustrates these contours for $Re=5\times 10^3$ at several streamwise locations. Close to the inflow, at $x/D=0.8$ and $1.2$, the contours exhibit a clear critical point structure in the upper half of the complex $\omega$-plane, indicating absolute instability of the local mean profile. As the streamwise position increases, the critical point moves toward smaller growth rate and eventually toward the stable half-plane, showing that the instability weakens downstream. This behavior is consistent with the physical picture of a near wake that is self-excited immediately behind the body, but gradually loses its ability to sustain global oscillations as the velocity deficit recovers.}

\begin{figure}
    \centering
    \includegraphics[width=0.7\linewidth]{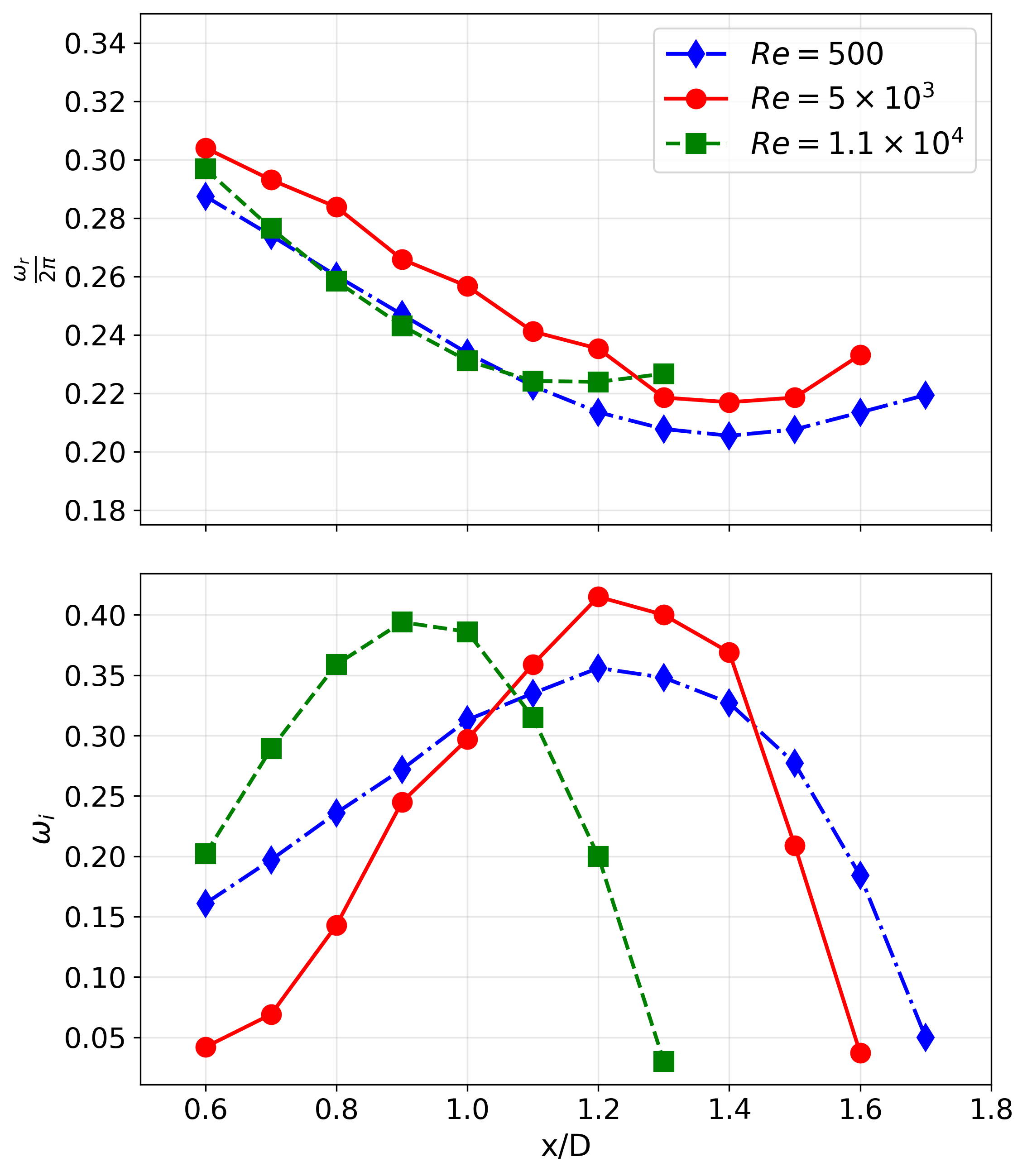}
    \caption{Critical frequency obtained from the local stability analysis as a function of streamwise location for $Re=500$, $5,000$, and $11,000$. Top: real part of the critical frequency, $\omega_r$. Bottom: imaginary part, $\omega_i$, which determines the local absolute or convective character of the wake. Positive values of $\omega_i$ indicate absolute instability and identify the near-wake region suitable for prescribing inflow profiles in the body-free simulation.Here, $\frac{\omega_r}/{2\pi}$ denotes the frequency associated with the corresponding mean profile, which is expected to coincide with the dominant shedding frequency of the reconstructed wake.}
    \label{fig:most_unstable_locations}
\end{figure}

\rev{The dependence of the critical frequency on streamwise position and Reynolds number is summarized in Figure~\ref{fig:most_unstable_locations}. For all three Reynolds numbers, the real part of the critical frequency varies only moderately over the unstable region, whereas the imaginary part decreases monotonically with downstream distance. The most amplified region is located in the near wake, approximately within $x/D\lesssim 1.5$, and the growth rate approaches zero farther downstream. Among the cases considered here, the lower-Reynolds-number wakes retain positive growth rates over a wider streamwise interval, while the $Re=11,000$ wake becomes weakly unstable and then stable earlier. These results support the central modeling assumption of the present work: if the reduced simulation is forced with a mean profile extracted within the absolutely unstable portion of the wake, the downstream dynamics can regenerate the unsteady vortex-shedding process without explicitly including the cylinder.}

\rev{Accordingly, the inflow locations adopted later in the body-free simulations are chosen from this near-wake region. The local stability analysis is not used here as a precise predictive tool for the nonlinear saturated state; rather, it serves as a physics-based criterion for selecting inflow profiles that retain the essential instability mechanism of the wake. In this sense, the analysis provides the dynamical link between the classical absolute-instability interpretation of bluff-body wakes and the body-free simulation framework developed in the following section.}

\section{Body-free Simulation Model for Flow Past a Cylinder}
\label{sec:reduced_model}
\begin{figure}
    \centering
    \includegraphics[width=0.75\linewidth]{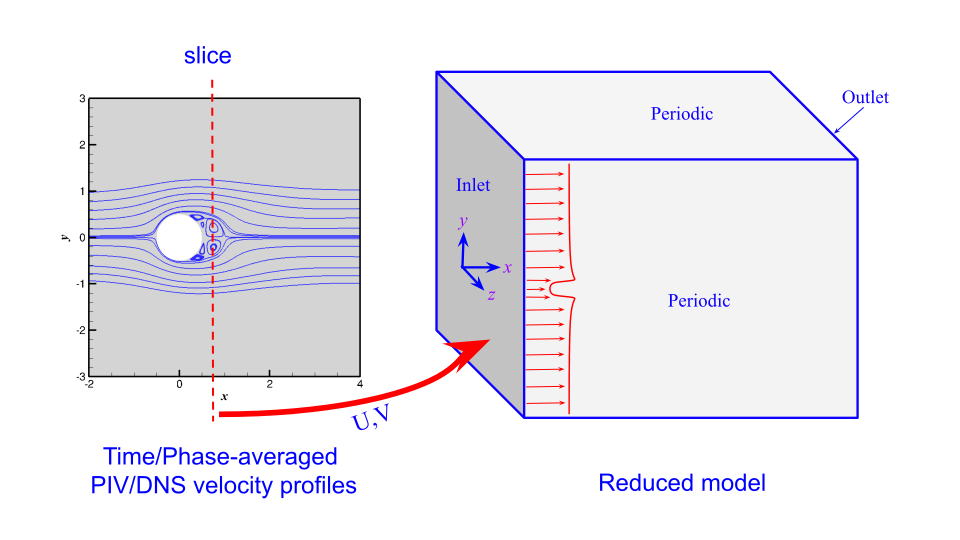}
    \caption{Computational setup for the body-free simulation. Unlike conventional direct numerical simulations (DNS) of cylinder flow, the present framework does not include a physical cylinder within the computational domain. Instead, one-dimensional velocity profiles representing the near wake---obtained from either experimental measurements or previous DNS with a physical cylinder---are prescribed at the inlet boundary.}
    \label{fig:reduced_model}
\end{figure}

In the body-free simulation, the three-dimensional incompressible Navier--Stokes equations are solved using a high-order spectral element method (SEM) \cite{GK_CFDbook} in a rectangular computational domain, as illustrated in Figure~\ref{fig:reduced_model}, without including a physical cylinder. \rev{A time-averaged or phase-averaged velocity profile, obtained for example from experiments or DNS, is prescribed at the inflow boundary without superimposing synthetic perturbations on the inlet profile itself.} To demonstrate the concept, we use both DNS data and experimental PIV data.
For the DNS-based inflow conditions, the phase-averaged field is computed by evenly sampling 32 snapshots over one vortex-shedding period and subsequently averaging over multiple periods. At the outflow boundary, homogeneous Neumann boundary conditions are applied to the velocity, while homogeneous Dirichlet conditions are imposed on the pressure. Periodic boundary conditions are imposed on the remaining four sides of the domain.

The computational domain has dimensions $30D$, $20D$, and $10D$ in the streamwise ($x$), crossflow ($y$), and spanwise ($z$) directions, respectively, where $D$ is the diameter of the imaginary circular cylinder. For the $Re=500$ case, the domain is discretized using $32\times32$ uniform quadrilateral elements in the $x$--$y$ plane, 256 Fourier planes in the $z$ direction, and spectral element polynomial order 4. For the higher-Reynolds-number cases, $Re=5{,}000$ and $Re=11{,}000$, the body-free simulations use $42\times42$ uniform quadrilateral elements in the $x$--$y$ plane and spectral element polynomial order 7. In addition, the $Re=5{,}000$ and $Re=11,000$ cases use 32 elements in the spanwise direction.

The body-free simulations were implemented using both our in-house SEM code \emph{Nektar2.5D} (\cite{NEWMAN_viv,wang2019jfm}), which employs Jacobi spectral element discretization in the $x$--$y$ plane and Fourier expansion in the $z$ direction, and the open-source SEM (Legendre) code \emph{nekRS} \cite{nekrs}. Specifically, the simulation at $Re=500$ is carried out using Nektar2.5D, whereas the simulations at the higher Reynolds numbers are conducted with nekRS, which can significantly accelerate computations on modern graphical processing units (GPUs).

\rev{All body-free simulations are initialized from a zero velocity field. To break the symmetry of the initial condition and accelerate the growth of unstable three-dimensional motions, a weak transient divergence-free volumetric forcing is applied during $0 \le tU_{\infty}/D \le 20$. The forcing is long-wavelength in all three spatial directions and is used only during start-up; its analytical form is given in Appendix~A. After the forcing is removed, the wake evolves freely until $tU_{\infty}/D=300$ to reach a statistically stationary state. Flow statistics are collected over $tU_{\infty}/D\in[200,300]$ for the lower-Reynolds-number case and over $tU_{\infty}/D\in[150,300]$ for the higher-Reynolds-number case.}

The experimental data were obtained from two-dimensional PIV measurements in the MIT Intelligent Towing Tank facility~\cite{FanPNAS2020}. A cylinder of diameter 25.4 mm was towed at a velocity of 0.188 m/s. Particle images in the $x$--$y$ plane were recorded using a high-speed camera and processed using DAVIS 11 with $32\times32$ pixel interrogation windows and 75\% overlap in the final pass. The data were non-dimensionalized using the cylinder diameter and towing velocity. The phase-averaged flow fields were obtained by applying proper orthogonal decomposition to the time series and then binning and averaging the snapshots according to their modal time coefficients, following Oudheusden et al. \cite{oudheusden2005phase}.
 
\section{Results and Discussion}

In the following, we present the body-free simulation results by first comparing the wake statistics with DNS and PIV data (Section~\ref{sec:results_statistics}) and then examining the impact of the inflow crossflow velocity on the three-dimensionality of the wake (Section~\ref{sec:results_cross-flow}).
\subsection{Statistics from the body-free simulation}
\label{sec:results_statistics}
\rev{In this section, we use time-averaged streamwise velocity profiles $\overline{U}(y)$ extracted from pre-computed DNS of flow past a cylinder as inflow boundary conditions for the body-free simulation. For the comparisons shown below, the inflow profile is taken at $x/D=0.75$ for the $Re=500$ and $Re=5{,}000$ cases, and at $x/D=0.9$ for the $Re=11{,}000$ case. These inflow locations are selected based on the local stability analysis as representative positions within the absolutely unstable near-wake region.}

\begin{figure}
  \centering
  \begin{subfigure}[t]{\textwidth}
  \centering
    \includegraphics[width=0.375\textwidth,trim=30 40 30 10,clip]{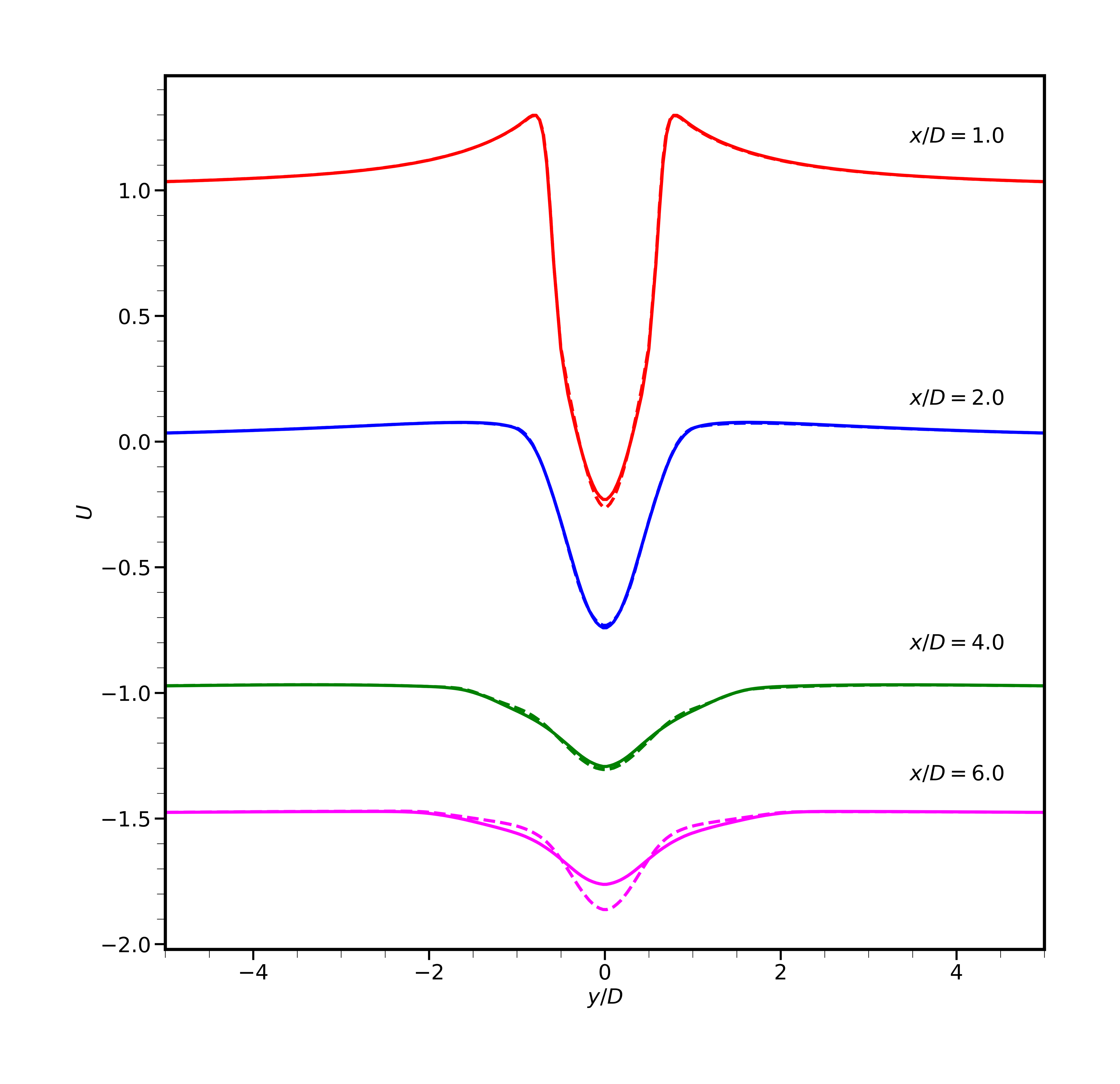}
        \hspace{5mm}
    \includegraphics[width=0.375\textwidth,trim=30 40 30 10,clip]{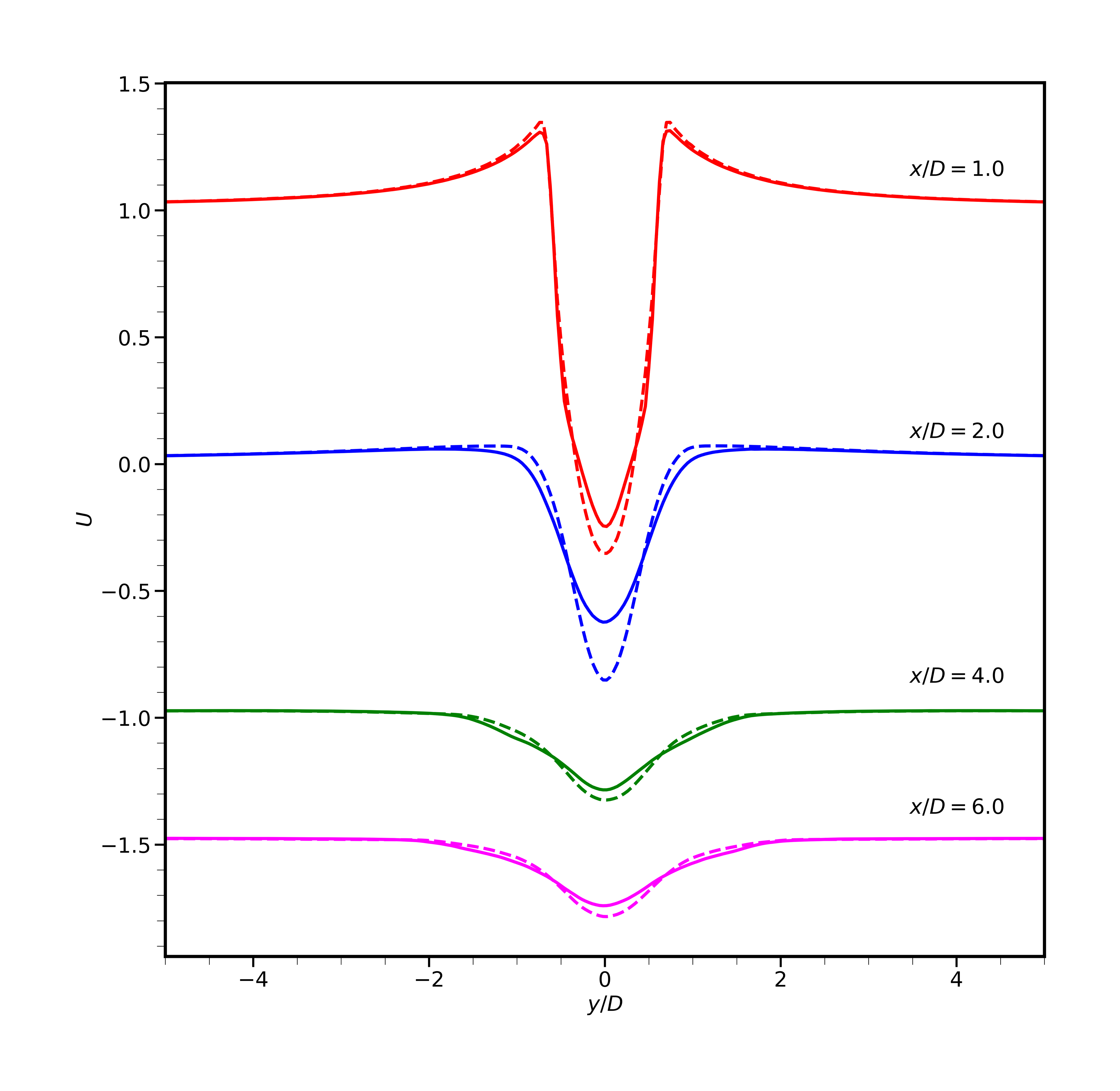}
        \label{fig:re5k_u}
      \vspace*{-5mm}
    \subcaption*{$\overline{U}$}
  \end{subfigure}

  \begin{subfigure}[t]{\textwidth}
  \centering
   \includegraphics[width=0.375\textwidth,trim=30 40 30 10,clip]{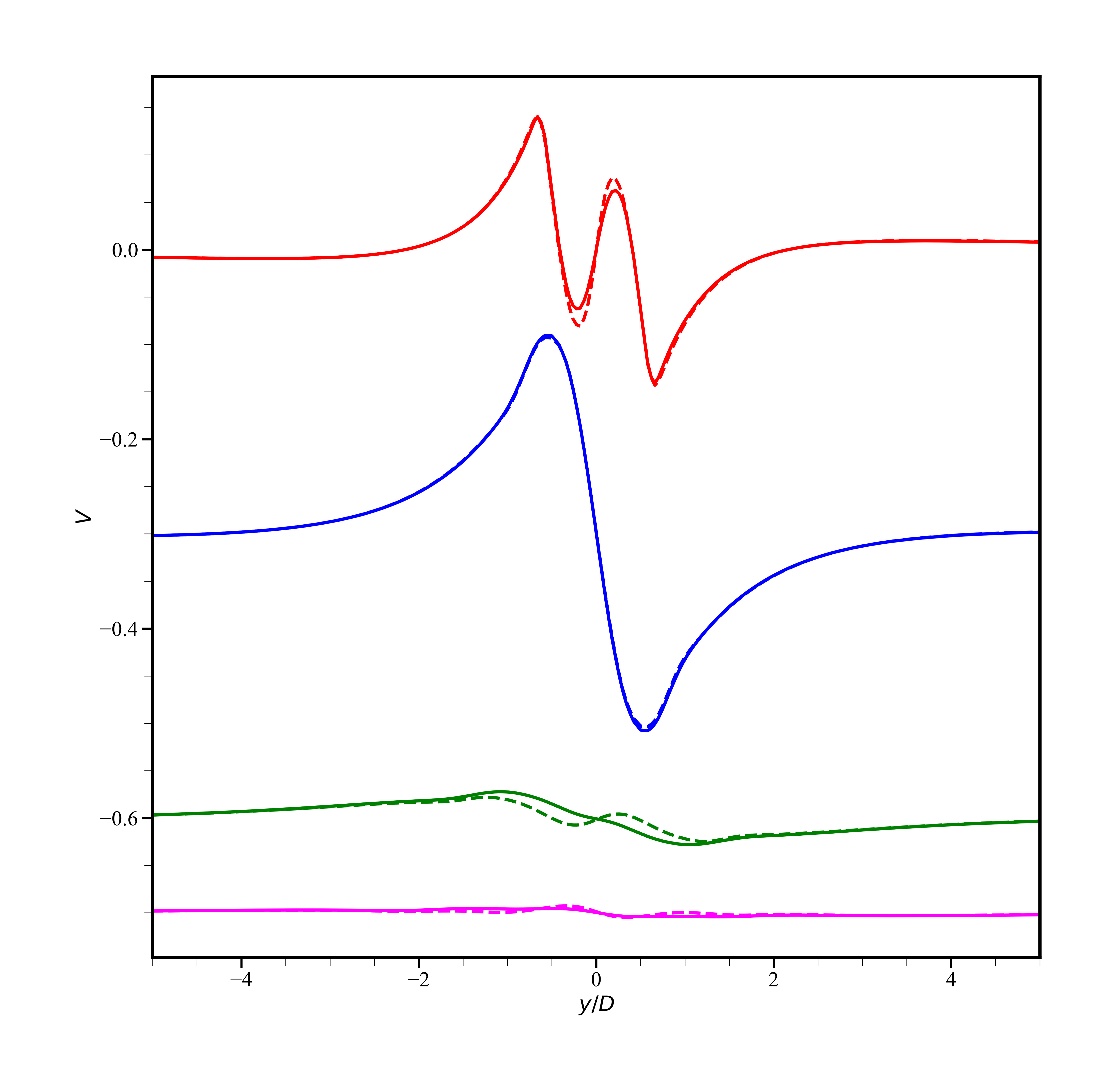}
       \hspace{5mm}
    \includegraphics[width=0.375\textwidth,trim=30 40 30 10,clip]{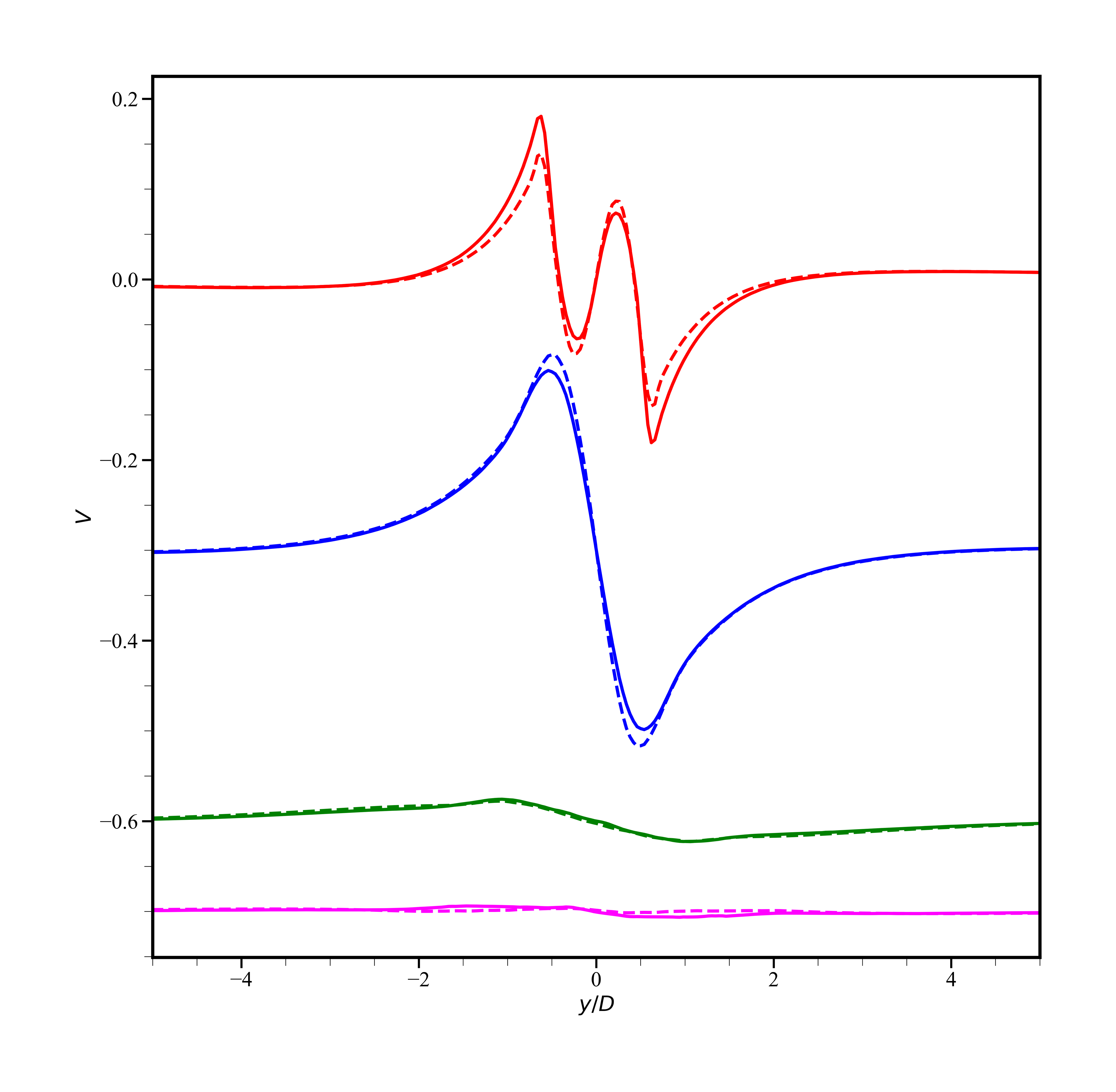}
    \label{fig:re5K_v}
      \vspace*{-5mm}
     \subcaption*{$\overline{V}$}
      \end{subfigure} 
~
   \caption{Comparison of the time-averaged velocity fields between DNS and the body-free simulation. Left: $Re=500$; right: $Re=5,\,000$. Solid lines denote DNS and dashed lines denote the body-free simulation. Red, blue, green, and magenta correspond to $x/D=1.0$, $2.0$, $4.0$, and $6.0$, respectively. Profiles other than $x/D=1.0$ are vertically shifted for clarity.}
    \label{fig:comparison_u}
\end{figure}

\begin{figure}
  \begin{subfigure}[t]{\textwidth}
  \centering
    \includegraphics[width=0.372\textwidth,trim=30 40 30 10,clip]{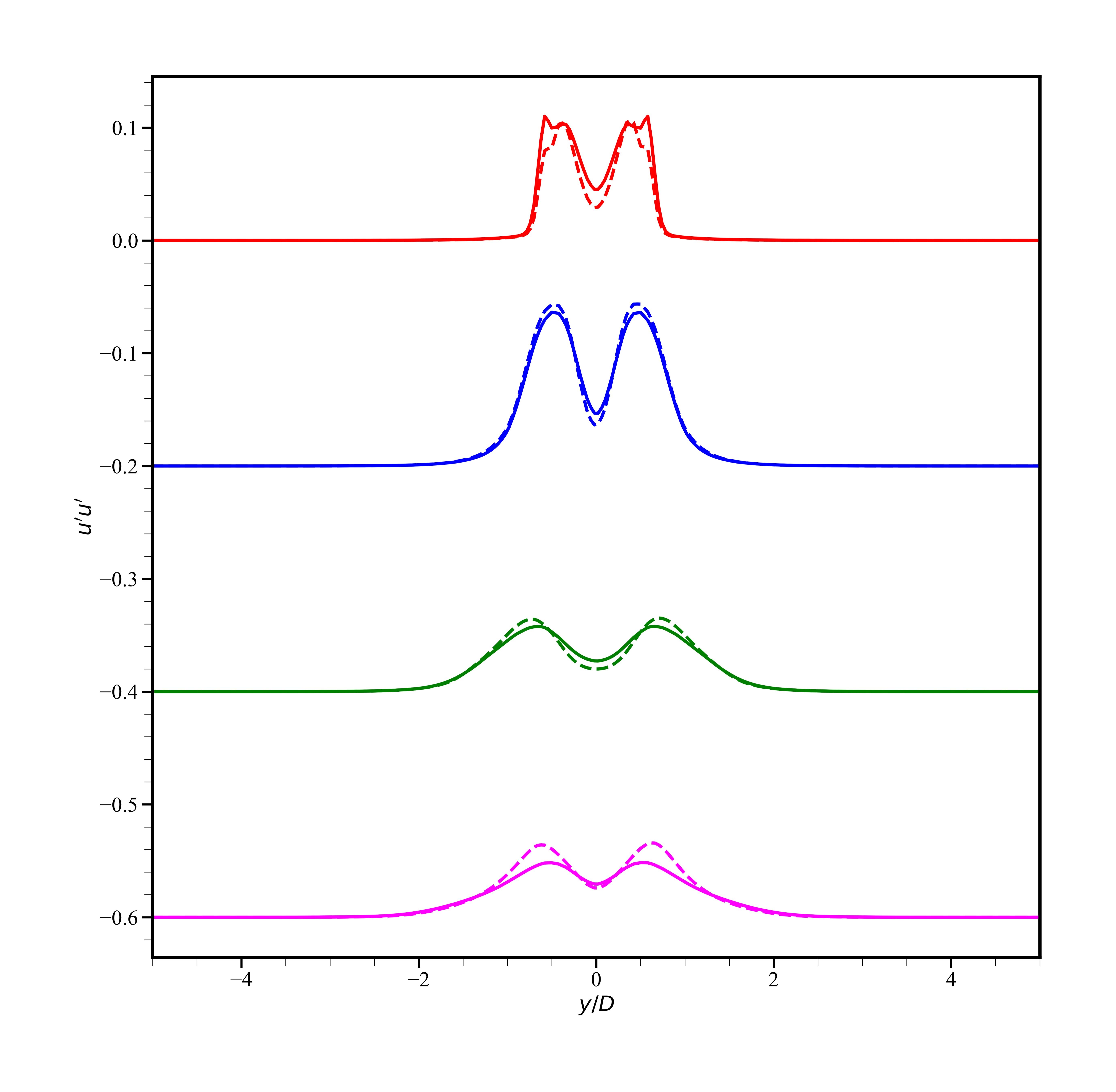}
    \hspace{5mm}
     \includegraphics[width=0.372\textwidth,trim=30 40 30 10,clip]{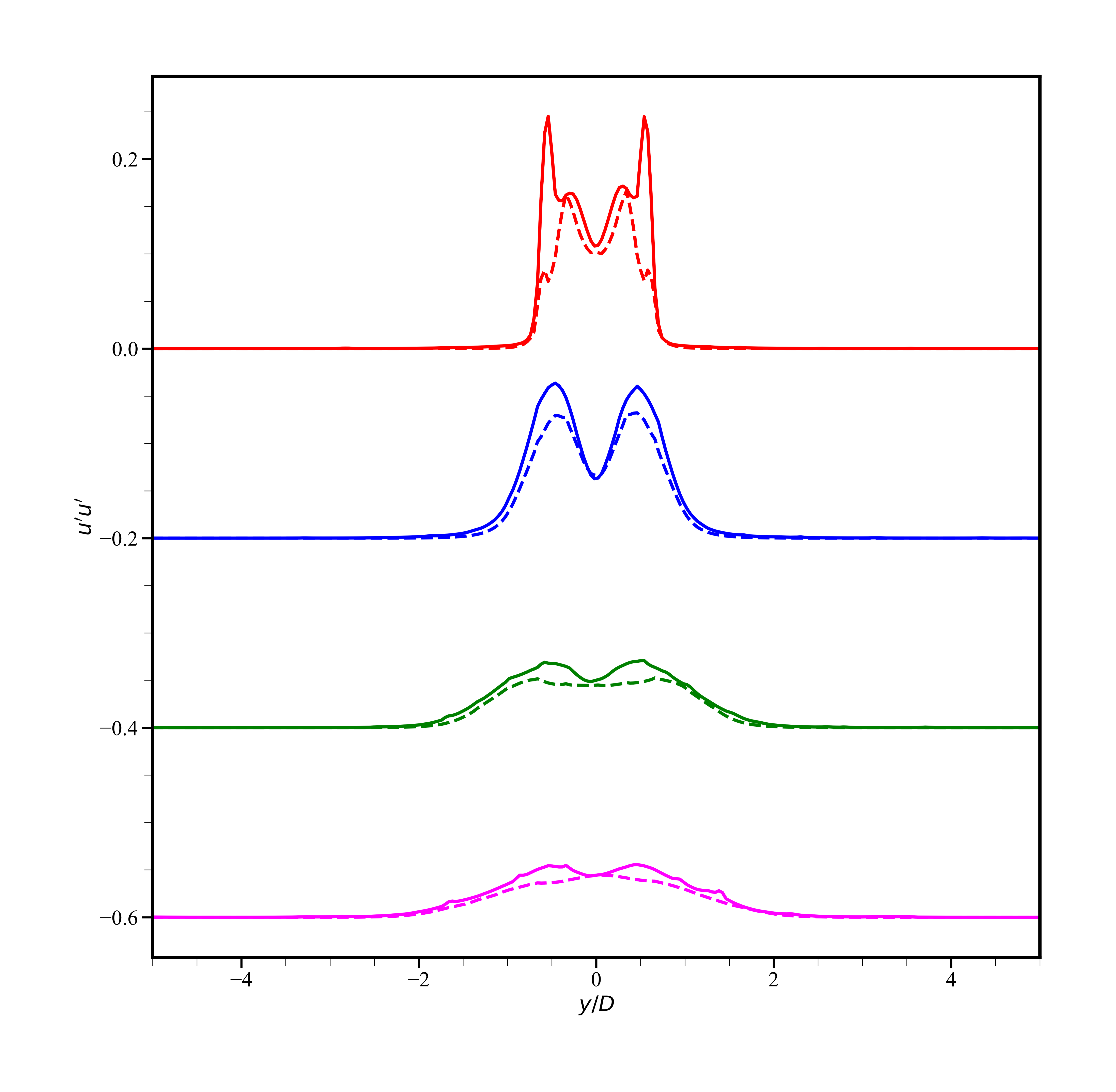}
        \label{fig:rek_uu}
      \vspace*{-5mm}
    \subcaption*{ $u^{\prime}u^{\prime}$ }
  \end{subfigure} 
  ~
    \begin{subfigure}[t]{\textwidth}
    \centering
    \includegraphics[width=0.372\textwidth,trim=30 40 30 10,clip]{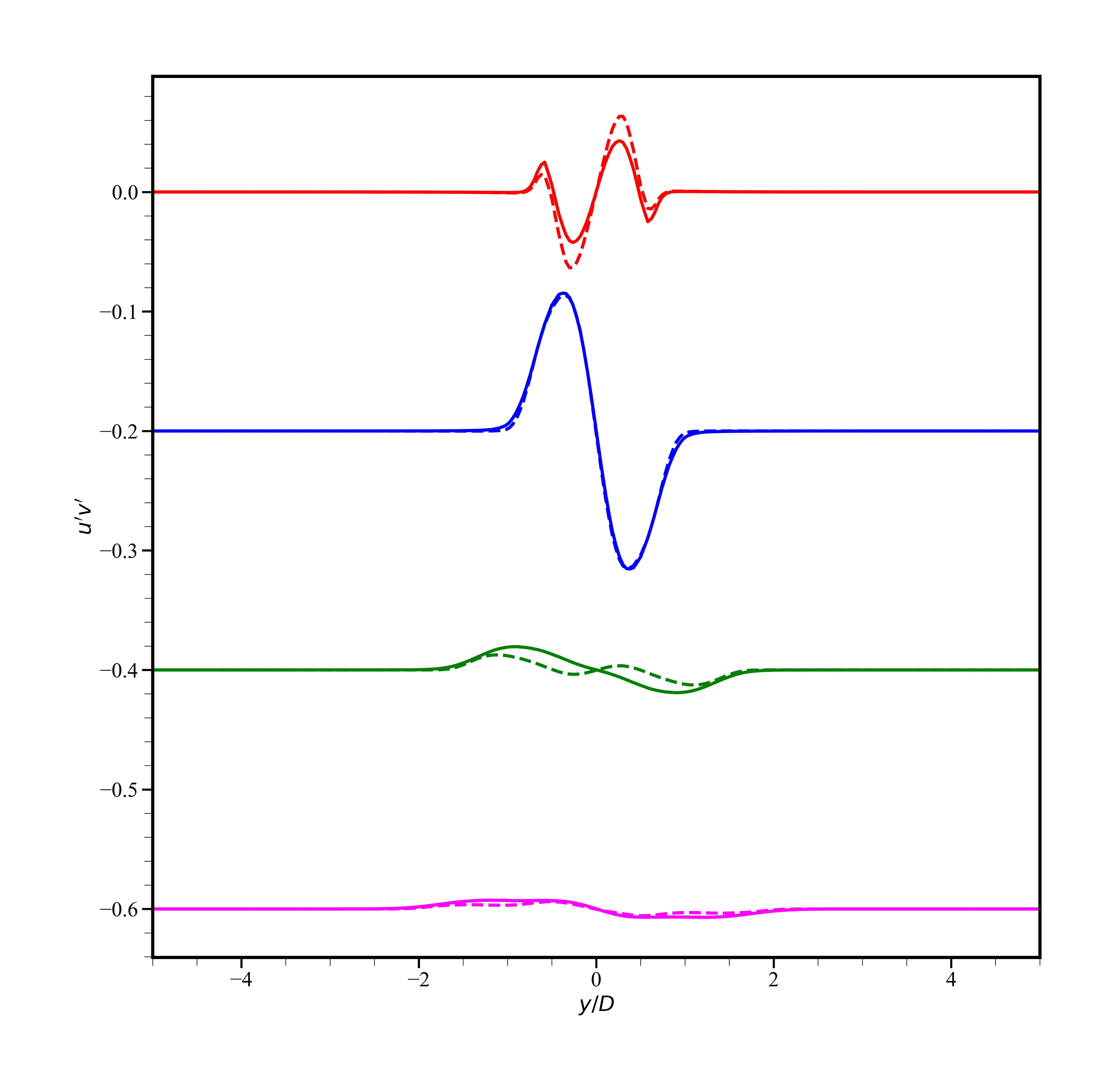}
        \hspace{5mm}
      \includegraphics[width=0.372\textwidth,trim=30 40 30 10,clip]{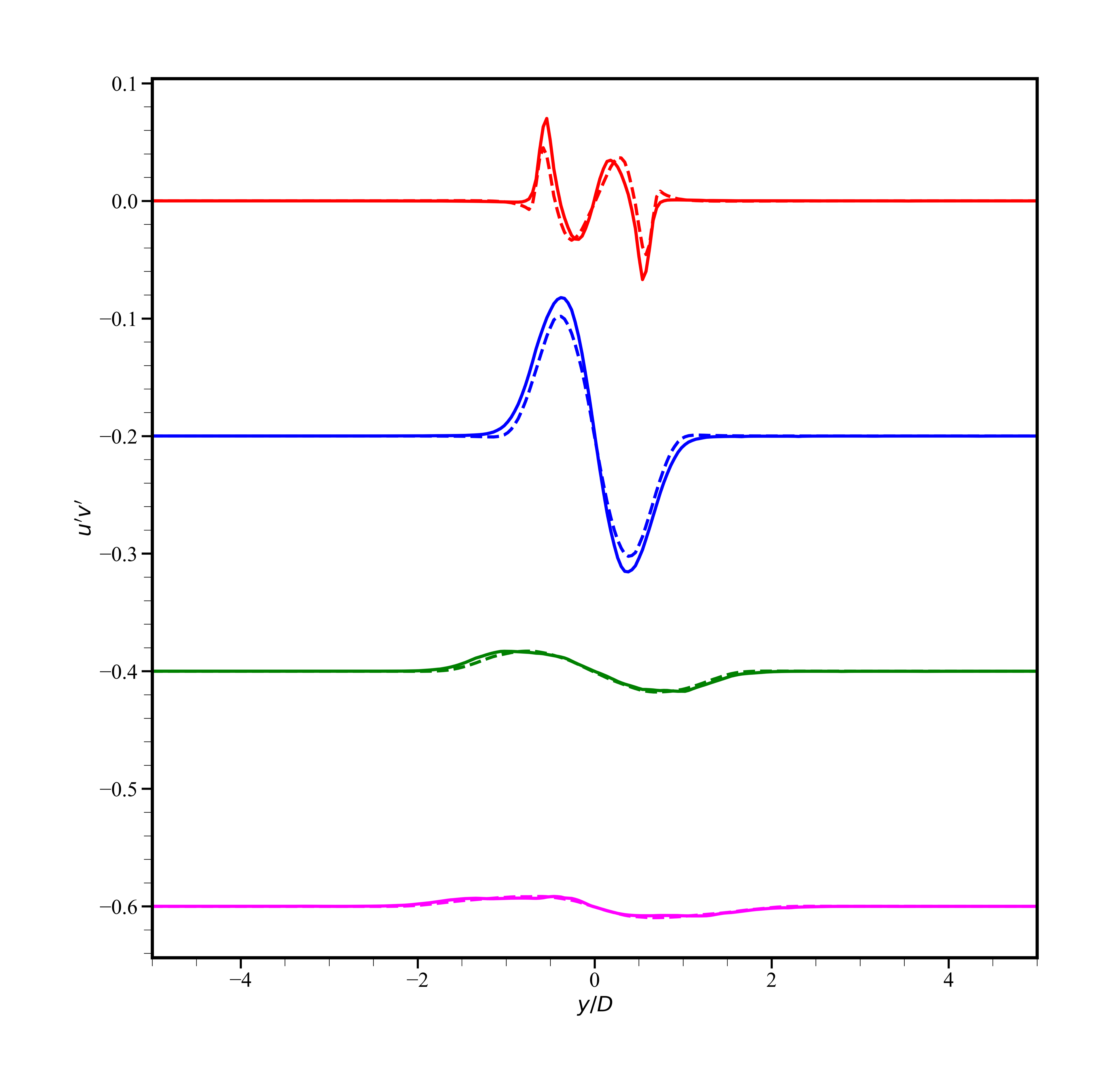}
        \label{fig:rek_uv}
      \vspace*{-5mm}
    \subcaption*{ $u^{\prime}v^{\prime}$ }
  \end{subfigure}    
   \caption{Comparison of the Reynolds stresses between DNS and the body-free simulation. Left: $Re=500$; right: $Re=5,\,000$. Profiles other than $x/D=1.0$ are vertically shifted for clarity. Line styles and colors follow Figure~\ref{fig:comparison_u}.}
   \vspace{-10pt}
       \label{fig:comparison_uu}
\end{figure}

When $\overline{U}$ is prescribed at an appropriate near-wake location, the body-free simulation shows good agreement with the corresponding full DNS with the cylinder present. Figures~\ref{fig:comparison_u} and \ref{fig:comparison_uu} compare time-averaged velocity profiles and Reynolds stresses at four downstream stations. At $Re=500$, the body-free simulation reproduces nearly all components of the mean flow and Reynolds stresses accurately, although small differences remain in the $v'v'$ profiles, for which the body-free simulation slightly overpredicts the values. At $Re=5{,}000$, the predicted $\overline{U}$ and $\overline{V}$ profiles also compare well with DNS. Moderate discrepancies appear in $\overline{U}$ at $x/D=2.0$, where the body-free simulation underpredicts the minimum value of the wake deficit. The predicted $u'u'$ profile in the near wake exhibits weaker peaks at $x/D=1.0$, which may be attributed to the absence of direct shear-layer forcing in the body-free simulation. The body-free simulation also captures $u'v'$ reasonably well. In addition, comparisons of $\overline{P}$, $v'v'$, and $w'w'$ (not shown) reveal moderate discrepancies in $v'v'$ and $w'w'$ at $x/D=2$ and some differences in $\overline{P}$ at $x/D=3$ for $Re=5{,}000$. The same body-free simulation setup used for the higher-Reynolds-number cases is then applied to $Re=11{,}000$, as discussed below.

~
~

\begin{figure}
\centering
  \begin{subfigure}[t]{0.372\textwidth}
    \includegraphics[width=\textwidth,trim=30 40 30 10,clip]{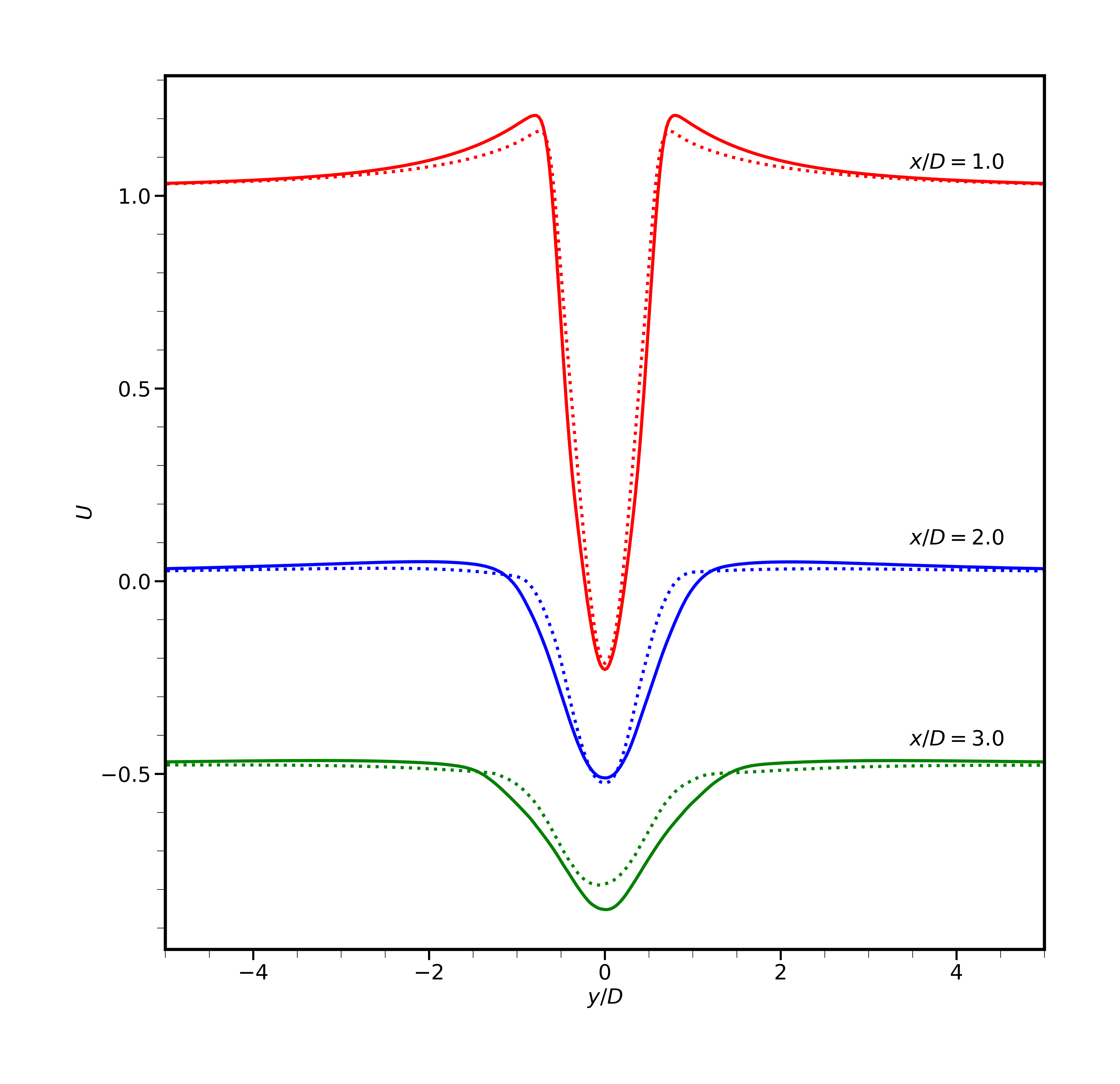}
        \label{fig:re11k_u}
      \vspace*{-5mm}
    \subcaption*{$\overline{U}$}
  \end{subfigure}
  \hspace{5mm}
  \begin{subfigure}[t]{0.372\textwidth}
    \includegraphics[width=\textwidth,trim=30 40 30 10,clip]{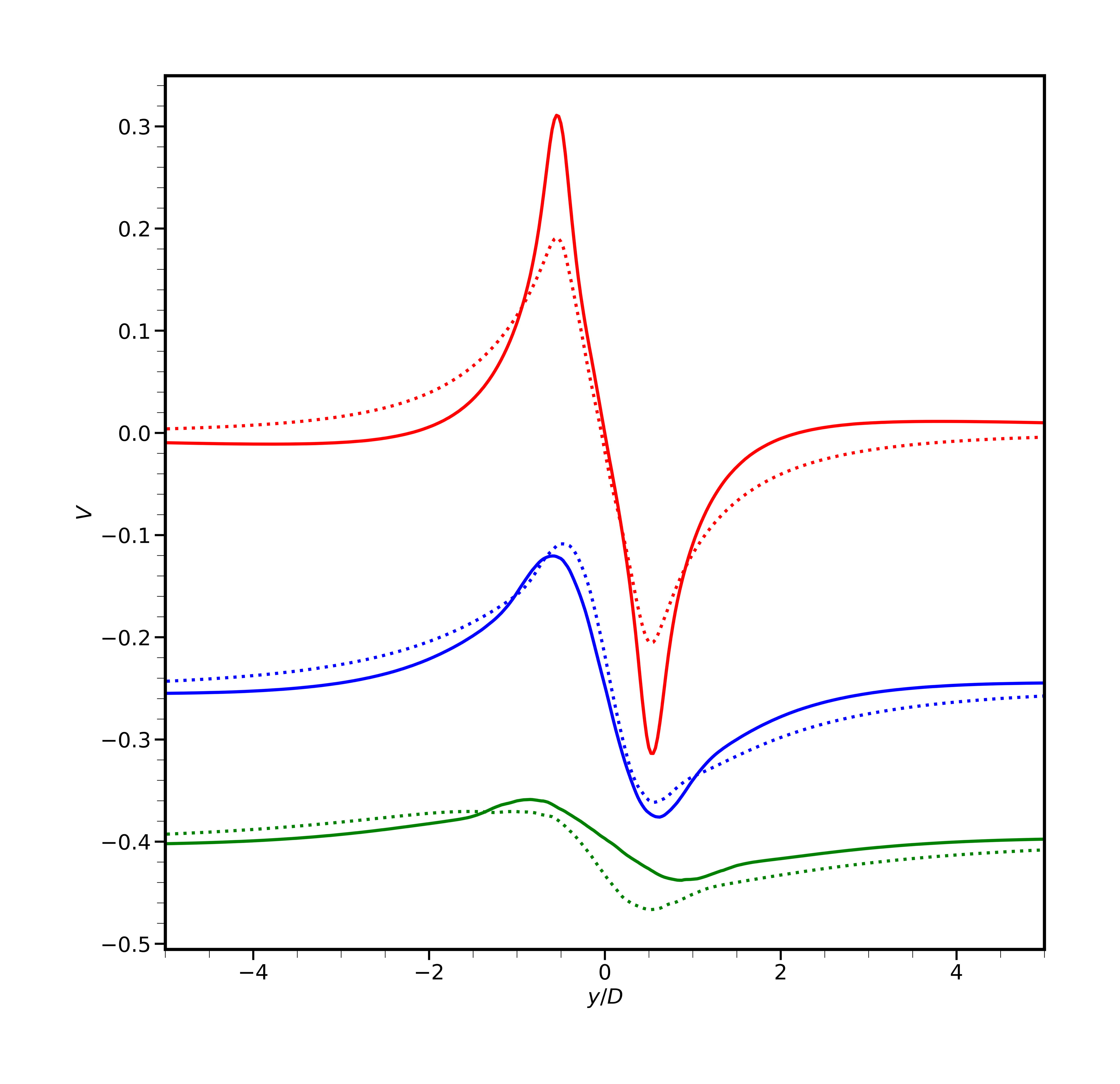}
    \label{fig:re11k_v}
      \vspace*{-5mm}
     \subcaption*{$\overline{V}$}
      \end{subfigure}   
       \begin{subfigure}[t]{0.375\textwidth}
    \includegraphics[width=\textwidth,trim=30 40 30 10,clip]{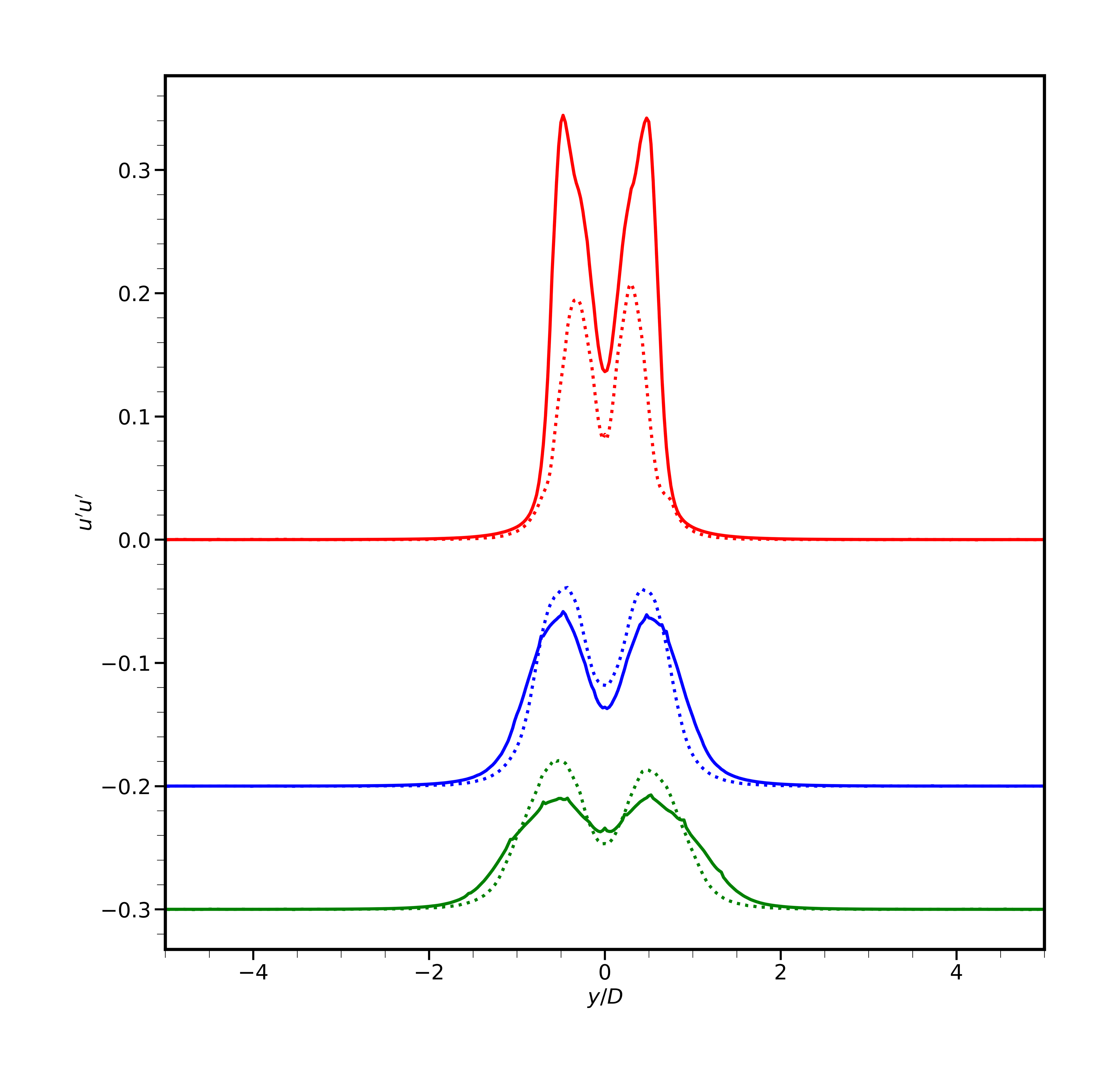}
        \label{fig:re11k_uu}
      \vspace*{-5mm}
    \subcaption*{$u^{\prime} u^{\prime}$}
  \end{subfigure}
  \hspace{5mm}
        \begin{subfigure}[t]{0.375\textwidth}
    \includegraphics[width=\textwidth,trim=30 40 30 10,clip]{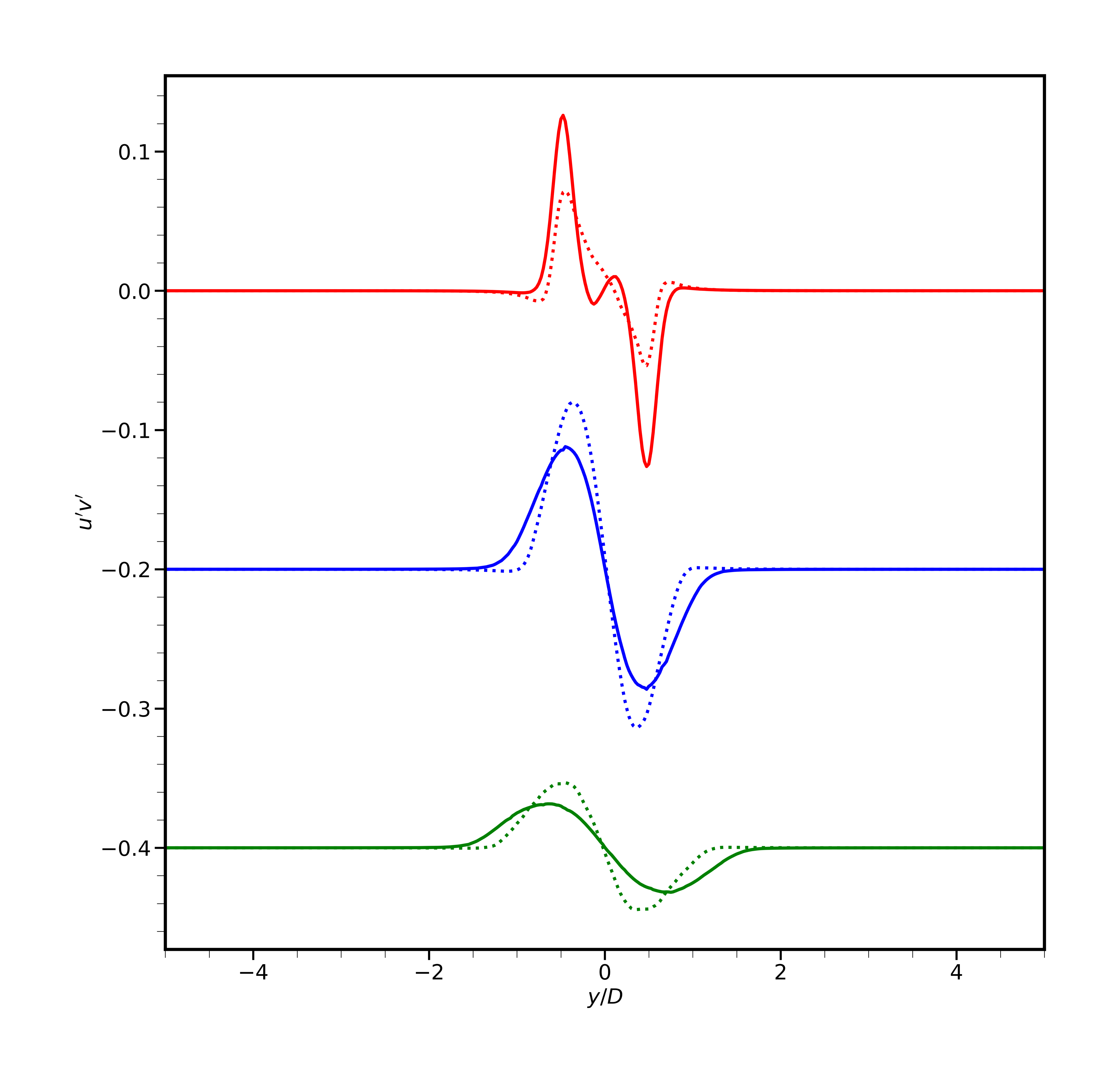}
    \label{fig:re11k_uv}
      \vspace*{-5mm}
     \subcaption*{$u^{\prime} v^{\prime}$}
      \end{subfigure}    
   \caption{Comparison of the time-averaged velocity fields and Reynolds stresses at $Re=11,\,000$ between DNS and the body-free simulation at several downstream locations in the wake. In the body-free simulation, the inflow boundary condition is prescribed using the time-averaged streamwise velocity extracted at $x/D=0.9$ from full DNS. Solid lines denote DNS and dashed lines denote the body-free simulation. Profiles other than $x/D=1.0$ are vertically shifted for clarity.}
       \label{fig:comparison_re11k_U}
\end{figure}

\rev{Figure~\ref{fig:comparison_re11k_U} extends the comparison to the higher-Reynolds-number case $Re=11{,}000$, for which the body-free simulation is driven only by the time-averaged streamwise velocity profile prescribed at $x/D=0.9$, identified by the stability analysis as the most unstable streamwise location. The body-free simulation again reproduces the overall structure of the wake with good agreement. The agreement is particularly encouraging for the mean velocity components $\overline{U}$ and $\overline{V}$, whose profiles at several downstream stations follow the DNS trends closely, including the wake deficit and its gradual recovery. The Reynolds-stress components are also captured reasonably well: the model reproduces the main shape and peak locations of $u'u'$ and $u'v'$, although some quantitative discrepancies remain in the near wake, where the body-free simulation smooths the strongest gradients and slightly underpredicts the peak amplitudes. Overall, these results indicate that the body-free simulation framework remains effective even at $Re=11{,}000$, suggesting that the dominant large-scale dynamics of the turbulent wake are still governed to a large extent by the prescribed near-wake mean profile.}

\rev{The Strouhal numbers reported in Table~\ref{tab:st_u_only} are consistent with the stability-analysis trends shown in Figure~\ref{fig:most_unstable_locations}. In particular, when the inflow profile is selected from the near-wake region identified as absolutely unstable by the local analysis, the body-free simulation predicts a sustained unsteady wake with a vortex-shedding frequency close to the critical frequency associated with that unstable region. This agreement is evident for $Re=500$ and $Re=5{,}000$. For $Re=11{,}000$, the case using $\overline{U}$ extracted at $x/D=0.9$ yields $St=0.247$, which is likewise consistent with the local stability analysis result $\frac{\omega_r}{2\pi}=0.243$, as shown in Figure~\ref{fig:most_unstable_locations}. These results support the assumption that the frequencies generated by the body-free simulation are selected by the instability characteristics of the prescribed mean profile.}

\begin{table}
\vspace{-10pt}
\caption{
Strouhal number ($St$) and wake type obtained from the body-free simulation when only the time-averaged streamwise velocity profile $\overline{U}$ is prescribed at the inflow. The inflow profiles are extracted from DNS at different downstream locations. Here, \textit{Type} denotes the nature of the wake, \textit{T} indicates a transitional wake, and ``None'' indicates that no distinct vortex shedding can be identified.
}
\begin{center}
\begin{adjustbox}{width=0.78\textwidth}
\centering
\begin{NiceTabular}{|c|c|c|c|c|c|c|c|c|}
\toprule
\multirow{3}{*}{$Re$} & \multirow{3}{*}{Quantity} & \multirow{3}{*}{DNS} & \multicolumn{6}{c}{Body-free simulation: time-averaged $\overline{U}$ only}\\
\cline{4-9}
& & & $x/D=0.75$ & $x/D=1.0$ & $x/D=1.2$ & $x/D=1.3$ & $x/D=1.4$ & $x/D=1.5$ \\
\hline
\multirow{2}{*}{$500$} & $St$ & 0.212 & 0.232 & 0.227 & 0.222 & 0.221 & 0.229 & 0.227\\
\cline{2-9}
& Type & 3D & 3D & 3D & 3D & 3D & 2D & 2D\\
\hline
\multirow{2}{*}{$5\,000$} & $St$ & 0.219 & 0.268 & 0.253 & 0.239 & 0.239 & 0.239 & 0.237\\
\cline{2-9}
& Type & 3D & 3D & 3D & 3D & 3D & 3D & 2D\\
\bottomrule
\end{NiceTabular}
\label{tab:st_u_only}
\end{adjustbox}
\end{center}
\end{table}

\subsection{Impact of Crossflow Velocity on Wake Behavior in the Body-Free Simulation}
\label{sec:results_cross-flow}
This subsection focuses on the role of the inflow crossflow component in the body-free simulation. In particular, we show that prescribing both phase-averaged velocity components at the inlet is important for recovering the correct vortex-shedding frequency, whereas changing the imposed crossflow profile can alter the wake type and even drive the flow toward a nearly two-dimensional state.

\begin{figure}
\centering
  \begin{subfigure}[t]{0.372\textwidth}
    \includegraphics[width=\textwidth,trim=30 40 30 10,clip]{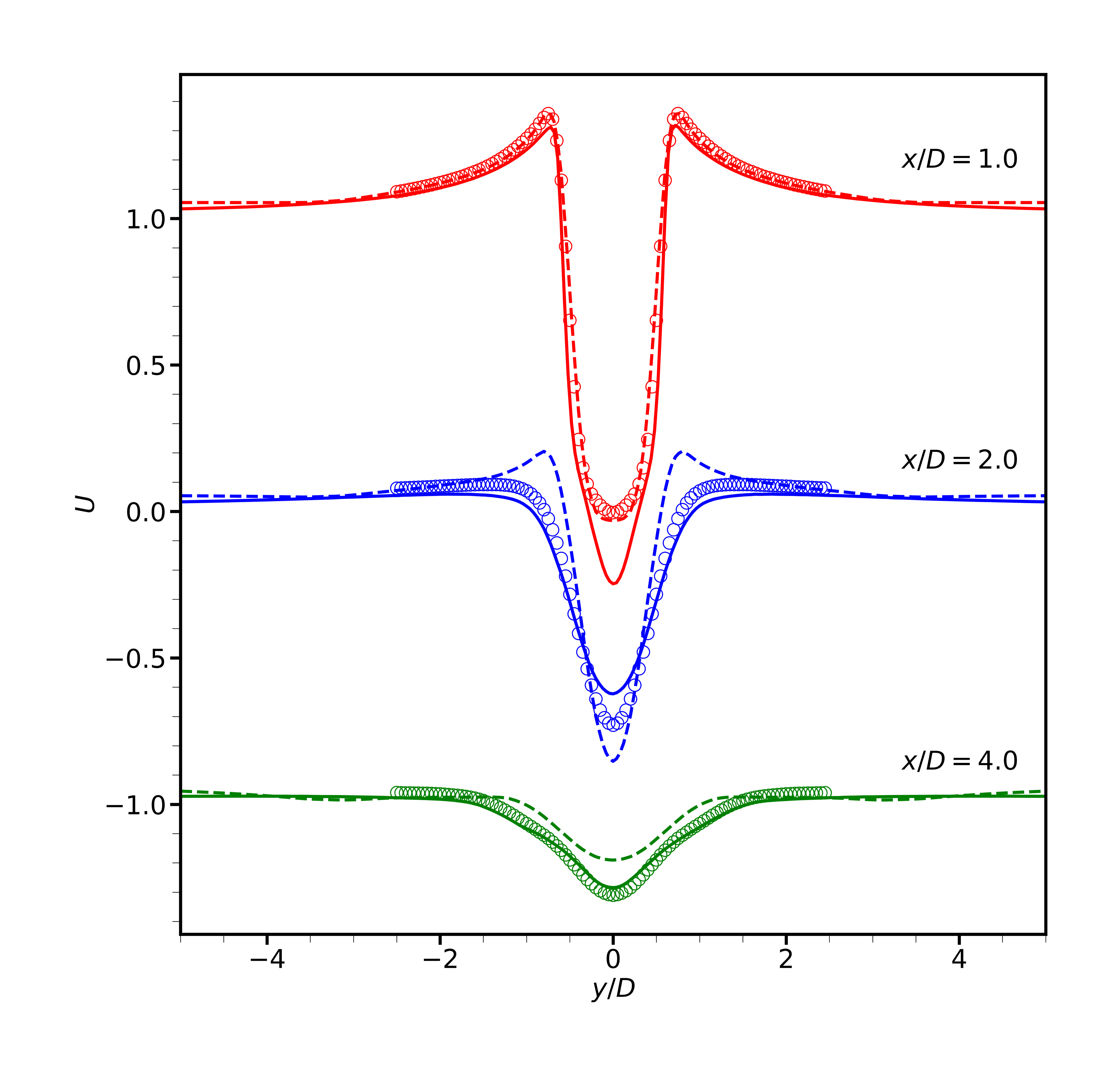}
        \label{fig:piv_re5k_u}
      \vspace*{-5mm}
    \subcaption*{$\overline{U}$}
  \end{subfigure}
  \hspace{5mm}
  \begin{subfigure}[t]{0.372\textwidth}
    \includegraphics[width=\textwidth,trim=30 40 30 10,clip]{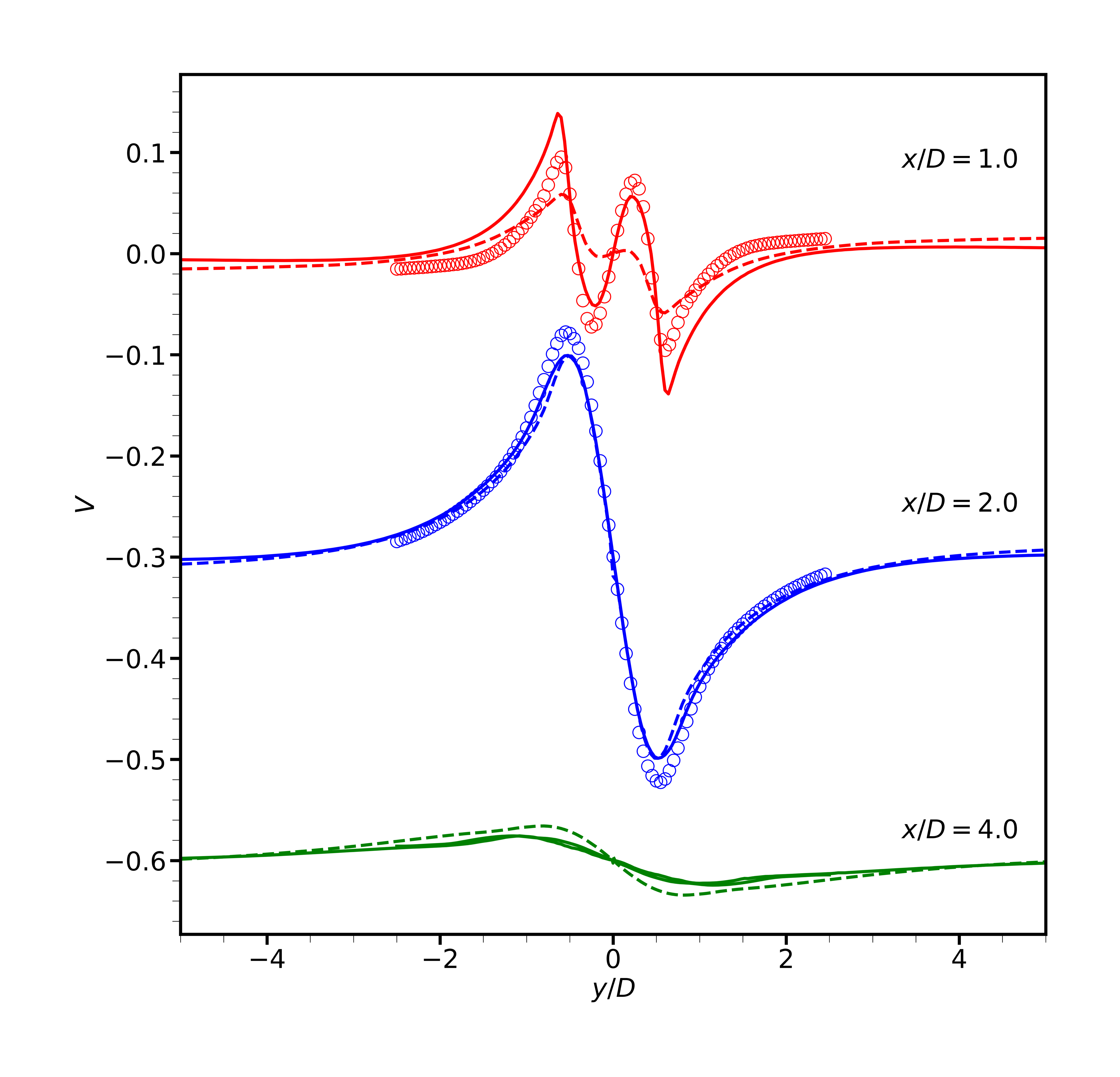}
    \label{fig:piv_re5k_v}
      \vspace*{-5mm}
     \subcaption*{$\overline{V}$}
      \end{subfigure}   
       \begin{subfigure}[t]{0.375\textwidth}
    \includegraphics[width=\textwidth,trim=30 40 30 10,clip]{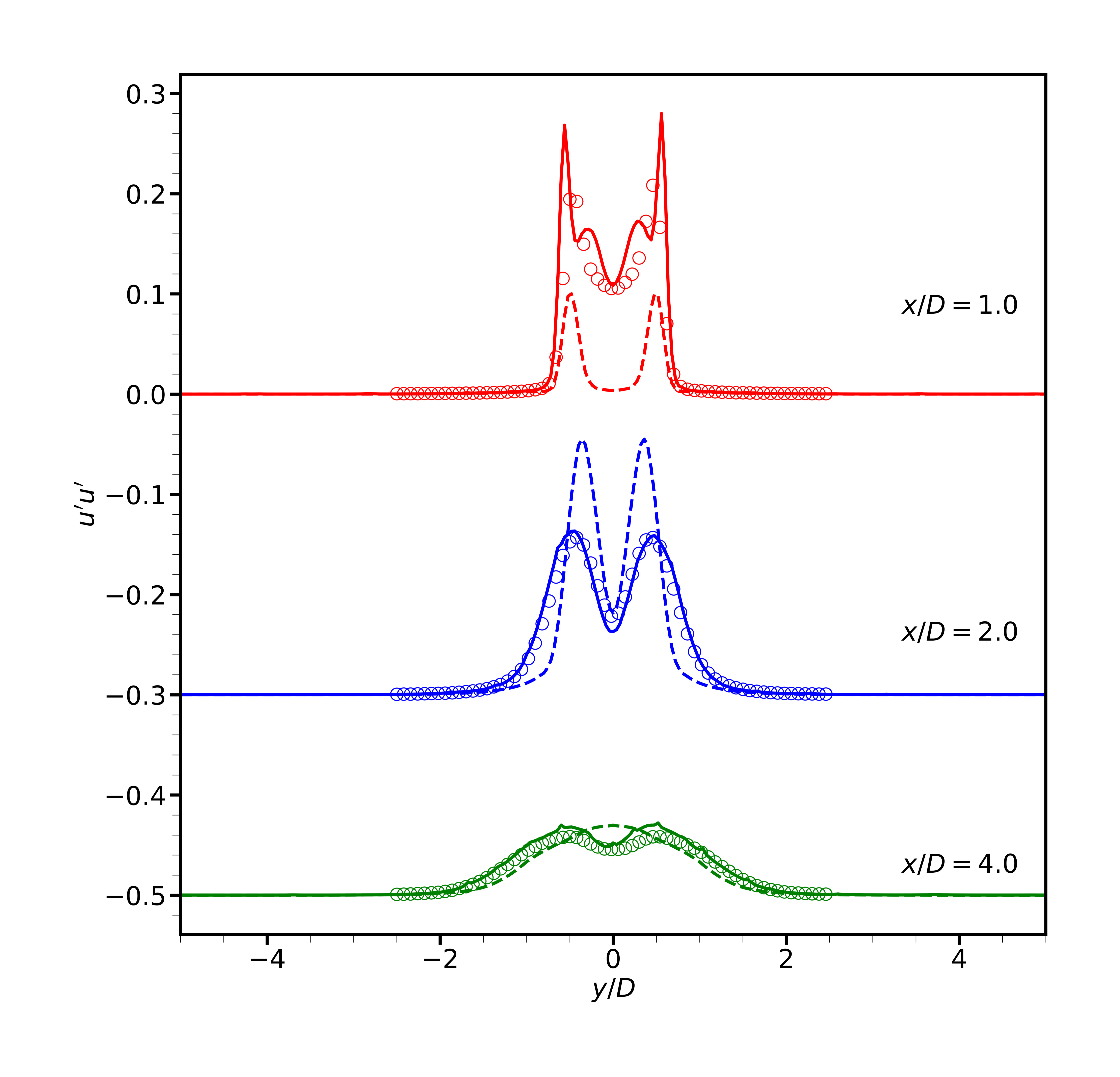}
        \label{fig:piv_re5k_uu}
      \vspace*{-5mm}
    \subcaption*{$u^{\prime} u^{\prime}$}
  \end{subfigure}
  \hspace{5mm}
        \begin{subfigure}[t]{0.375\textwidth}
    \includegraphics[width=\textwidth,trim=30 40 30 10,clip]{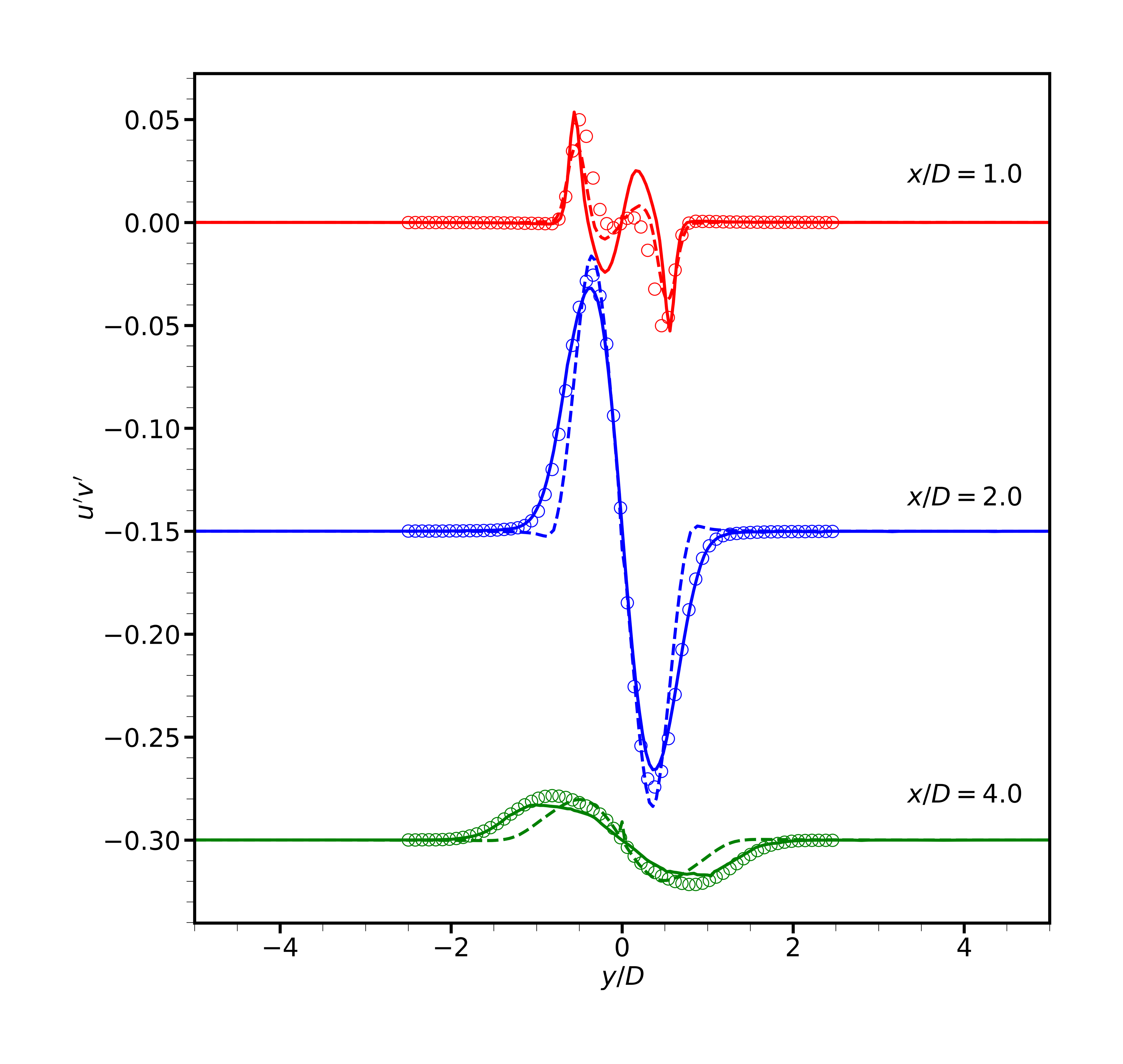}
    \label{fig:piv_re5k_uv}
      \vspace*{-5mm}
     \subcaption*{$u^{\prime} v^{\prime}$}
      \end{subfigure}    
   \caption{Comparison of the time-averaged velocity fields and Reynolds stresses at $Re=5,\,000$ between the PIV measurements and the body-free simulation results at several downstream locations in the wake. In the body-free simulation, the inflow boundary condition is prescribed using the phase-averaged streamwise and crossflow velocity components extracted at $x/D=0.6$ from the two-dimensional PIV measurements. Empty circles denote the PIV data, solid lines denote DNS, and dashed lines denote the body-free simulation. Profiles other than $x/D=1.0$ are vertically shifted for clarity.}
       \label{fig:comparison_piv_re5k_U}
\end{figure}

We first present the PIV-based results. The inflow boundary condition is prescribed using the phase-averaged streamwise and crossflow velocity components extracted at $x/D=0.6$ from two-dimensional PIV measurements. The PIV measurement window spans $[-0.55D,8.5D] \times [-4.0D,4.0D]$ in the streamwise and crossflow directions, respectively. In the body-free simulation, the inflow boundary condition is imposed over $-3.5 \le y/D \le 3.5$ using the measured data at $x/D=0.6$; outside this interval, both velocity components are held constant in $y$. Figure~\ref{fig:comparison_piv_re5k_U} compares the mean streamwise and crossflow velocity components, as well as the Reynolds stresses, between the PIV data and the body-free simulation. At $x/D=1$, the body-free simulation reproduces the $\overline{U}$ profile accurately. At $x/D=2$, it predicts a somewhat stronger velocity deficit, while at $x/D=4$ the wake profile is slightly shallower than in the PIV measurements. The top-right panel of Figure~\ref{fig:comparison_piv_re5k_U} shows that the body-free simulation underpredicts $\overline{V}$ at $x/D=1.0$, whereas the agreement improves farther downstream. The lower-left panel shows that the body-free simulation generally underpredicts the Reynolds stress component $u'u'$ relative to the PIV data, except at $x/D=2.0$, where it exceeds the PIV value at the same location where $\overline{U}$ exhibits the strongest deficit. By contrast, the agreement in $u'v'$ is noticeably better. The corresponding DNS results are also included in Figure~\ref{fig:comparison_piv_re5k_U} for reference. Overall, the body-free predictions are closer to the PIV data than to the DNS, as expected, because the inflow boundary condition is derived directly from the PIV measurements.

To quantify the role of the inflow specification, Table~\ref{tab:st_u_only} summarizes the wake type and the corresponding Strouhal number obtained when only the time-averaged streamwise velocity $\overline{U}$ is prescribed at the inlet, whereas Table~\ref{tab:st_with_v_phase} collects the results obtained when either phase-averaged inflow profiles or both time-averaged velocity components $(\overline{U},\overline{V})$ are imposed. A clear trend emerges: for both $Re=500$ and $Re=5{,}000$, the body-free simulation recovers the correct vortex-shedding frequency only when the inflow includes both velocity components in phase-averaged form. In contrast, using only time-averaged $\overline{U}$, or using time-averaged $(\overline{U},\overline{V})$, can still generate an unsteady wake, but the resulting $St$ is shifted away from the DNS or PIV value.

The tables also show that the crossflow velocity strongly affects the wake topology. When only $\overline{U}$ is prescribed, the body-free simulation can sustain a three-dimensional wake over a relatively broad range of extraction locations. However, once a time-averaged $\overline{V}$ profile is added, the admissible range becomes narrower, and the wake can transition from three-dimensional to transitional or nearly two-dimensional behavior. This sensitivity indicates that the streamwise velocity deficit supplies the energy required for wake amplification, whereas the imposed crossflow profile influences how that instability saturates. In particular, the phase-averaged crossflow component provides the appropriate transverse forcing needed to recover the physical shedding cycle, while time-averaged $\overline{V}$ profiles extracted from unsuitable locations can suppress three-dimensionality and alter the wake type.

\rev{This distinction is especially important from an experimental perspective. In numerical simulations, one may prescribe the full inflow velocity vector or even a phase-averaged inflow profile relatively easily. In a physical experiment, however, it is generally much more difficult to impose a spatially varying crossflow component $V$, and it is even less practical to enforce a phase-averaged inflow synchronized with the vortex-shedding cycle. In many realistic experimental settings, the quantity that is most readily available is the mean streamwise velocity profile $\overline{U}$. From this standpoint, it is encouraging that the body-free simulation can still recover a sustained three-dimensional wake using only $\overline{U}$, even though the prediction of the Strouhal number improves when additional crossflow or phase information is supplied. Thus, the results suggest that phase-averaged inflow data provide the most accurate wake reconstruction, but mean streamwise inflow alone may already be sufficient for many experimentally relevant reduced-complexity applications.}

\begin{table}
\vspace{-10pt}
\caption{
Strouhal number ($St$) and wake type obtained from the body-free simulation using either phase-averaged inflow profiles $(\widetilde{U},\widetilde{V})$ or both time-averaged velocity components $(\overline{U},\overline{V})$. All inflow profiles are extracted from DNS. Here, \textit{Type} denotes the nature of the wake, \textit{T} indicates a transitional wake, and ``None'' indicates that no distinct vortex shedding can be identified.
}

\begin{center}
\begin{adjustbox}{width=0.9\textwidth}
\centering
\begin{NiceTabular}{|c|c|c|c|c|c|c|c|c|c|}
\toprule
\multirow{2}{*}{$Re$} & \multirow{2}{*}{Quantity} &  \multirow{2}{*}{DNS} & \multirow{2}{*}{Phase-averaged $(\widetilde{U},\widetilde{V})$} & \multicolumn{6}{c}{Time-averaged $(\overline{U},\overline{V})$}\\
\cline{5-10}
& &  &  & $x/D=0.75$ & $x/D=1.0$ & $x/D=1.2$ & $x/D=1.3$ & $x/D=1.4$ & $x/D=1.5$ \\
\hline
\multirow{2}{*}{$500$} & $St$ & 0.212 & 0.212 & 0.234 & 0.235 & 0.265 & None & None & None\\
\cline{2-10}
& Type & 3D & 3D & 3D & 3D & 2D & None & None & None\\
\hline
\multirow{2}{*}{$5\,000$} & $St$ & 0.219 & 0.220 & 0.268 & 0.252 & 0.305 & 0.299 & None & None\\
\cline{2-10}
& Type & 3D & 3D & 3D & 3D & 3D & T & None & None\\
\bottomrule
\end{NiceTabular}
\label{tab:st_with_v_phase}
\end{adjustbox}
\end{center}
\end{table}

\rev{These results suggest that the prescribed $\overline{U}$ profile largely determines whether the instability contains sufficient energy to develop into a three-dimensional wake. For example, at $Re=5{,}000$, the body-free simulation predicts a three-dimensional wake as long as $\overline{U}$ is extracted from a location within $x/D\le 1.4$. However, when the corresponding time-averaged $\overline{V}$ profile extracted from the same location is also imposed at the inflow boundary, the wake begins to transition toward a more two-dimensional state. As shown in the lower panel of Figure~\ref{fig:comparison_u}, the near-wake $\overline{V}$ profile exhibits two characteristic shapes: (i) a four-peak antisymmetric structure within the recirculation bubble and (ii) a two-peak antisymmetric structure farther downstream. Because the peaks in the second structure are stronger, their imposition at the inlet tends to promote the transition from a three-dimensional to a more two-dimensional wake.}

\rev{Before presenting the crossflow-scaling results, we clarify their interpretation. The cases using $(\overline{U},\alpha\overline{V})$ are intended as sensitivity studies rather than exact reconstructions of a parent wake. Once $\overline{V}$ is rescaled independently of $\overline{U}$, the imposed inlet pair should be interpreted as a controlled modification of the boundary forcing. The purpose is to examine how the crossflow content of the imposed near-wake profile affects frequency selection, wake saturation, and the emergence or suppression of three-dimensionality. Although the modified inlet data do not by themselves represent a fully self-consistent extracted plane, the computed flow remains divergence-free because incompressibility is enforced by the pressure-projection step of the Navier--Stokes solver.}

\begin{figure}
    \centering
    \includegraphics[width=0.7\linewidth]{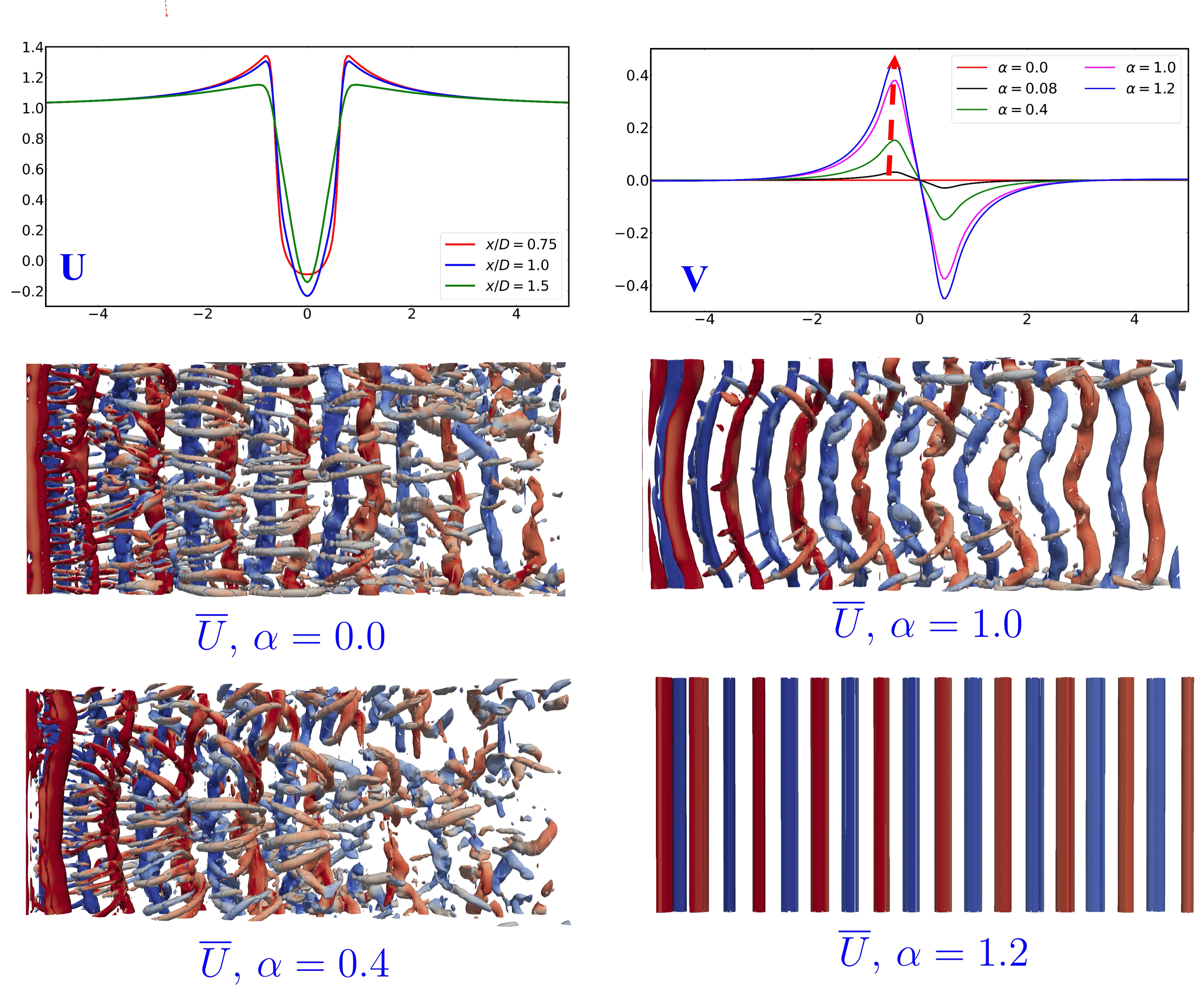}
\caption{\rev{Wake structures from the body-free simulation using different inflow velocity profiles at $Re=500$. The top row shows the time-averaged streamwise and crossflow velocity profiles extracted from DNS of flow past a cylinder. The remaining panels show iso-surfaces of $Q=0.1$, colored by the spanwise vorticity component $\omega_z$ in the range $[-2,2]$. In these simulations, the inflow streamwise component is fixed to the time-averaged profile $\overline{U}$ extracted at $x/D=0.75$, whereas the inflow crossflow component is taken as $V=\alpha\overline{V}$, where $\overline{V}$ is the time-averaged profile extracted at $x/D=1.4$. Thus, $\alpha$ parametrizes a controlled modification of the imposed crossflow content at the inlet; for example, $\alpha=0.4$ corresponds to $V=0.4\overline{V}$. Because of the absolute instability of the wake \cite{Triantafyllou1986Formation}, the body-free simulation with $\alpha=0$ can still produce a von K\'arm\'an street. As $\alpha$ increases, the wake vortices weaken and the flow eventually becomes two-dimensional.}}
\vspace{-3mm}
\label{fig:Re500_trans}
\end{figure}

\rev{To investigate the influence of the inflow crossflow component more systematically for $Re=500$, we select the time-averaged $\overline{U}$ profile extracted at $x/D=0.75$ and the time-averaged $\overline{V}$ profile extracted at $x/D=1.4$, corresponding to the red line in the top-left panel and the magenta line in the top-right panel of Figure~\ref{fig:Re500_trans}. We then perform a sensitivity study in which the imposed crossflow component is varied parametrically by replacing $\overline{V}$ with $\alpha\overline{V}$, while keeping the imposed streamwise profile fixed. Here, $\alpha=0$ corresponds to the case $V=0$, while $\alpha=1.2$ corresponds to an inflow crossflow component equal to $1.2\overline{V}$. Figure~\ref{fig:Re500_trans} shows that when $\alpha=0$, that is, when only $\overline{U}$ is imposed at the inlet, the vortical structures closely resemble those observed in DNS. As $\alpha$ increases, the braid vortices generated by the body-free simulation become progressively weaker, and at $\alpha=1.2$ the wake transitions from a three-dimensional to a two-dimensional state.}

 \begin{figure}
  \centering
  \begin{subfigure}[t]{0.472\textwidth}
    \includegraphics[width=\textwidth,trim=30 40 30 10,clip]{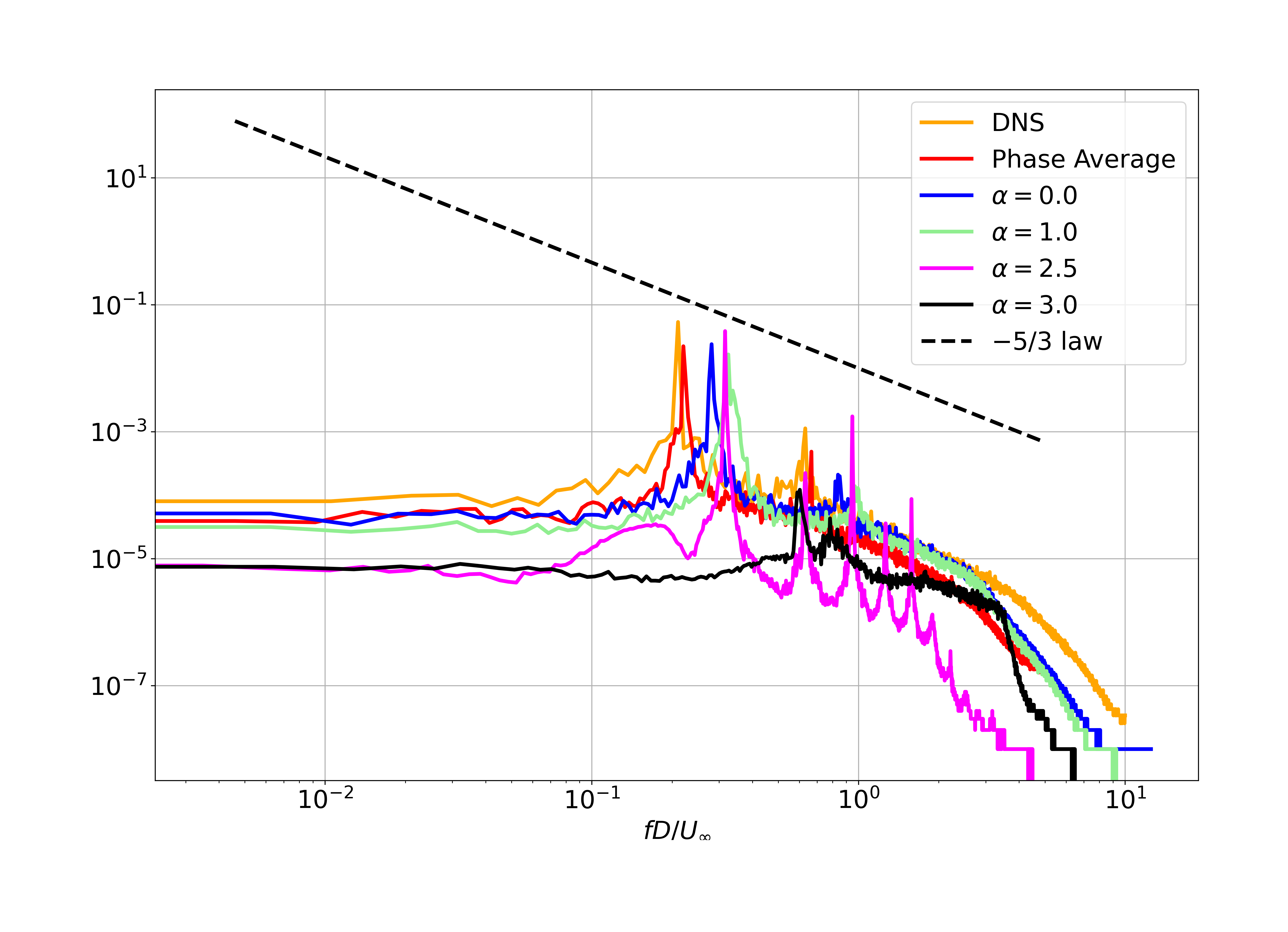}
        \label{fig:spectra_freq}
      \vspace*{-5mm}
    \subcaption*{Frequency spectra}
  \end{subfigure}
  \hspace{5mm}
  \begin{subfigure}[t]{0.472\textwidth}
    \includegraphics[width=\textwidth,trim=30 40 30 10,clip]{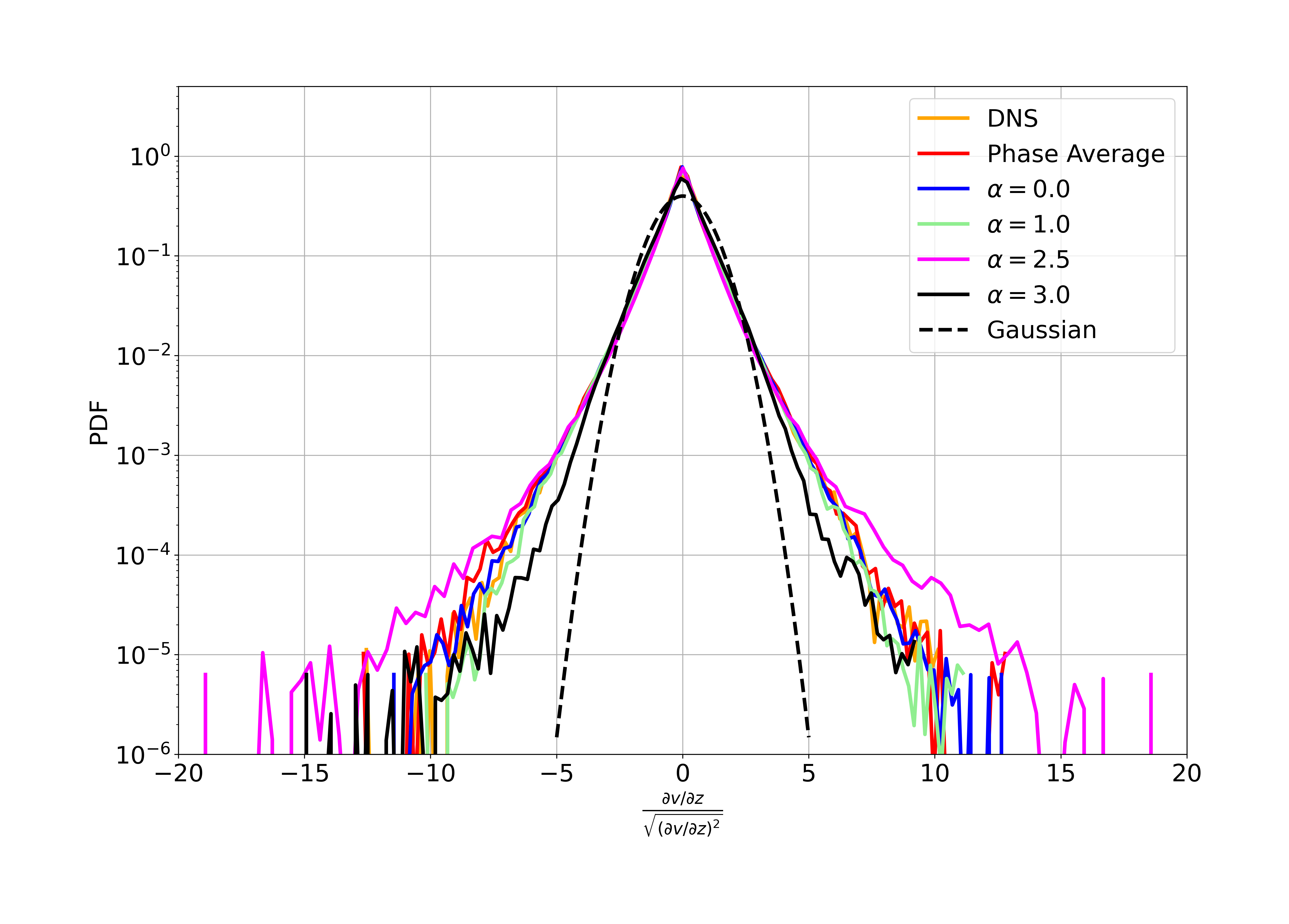}
    \label{fig:spectra_z}
      \vspace*{-5mm}
     \subcaption*{ Probability density function (PDF)}
      \end{subfigure}      
   \caption{One-dimensional spectra of the crossflow velocity $v$ and probability density functions of the spanwise velocity gradient $\partial v/\partial z$, obtained from data extracted along the line $x/D=6.05$, $y/D=0$, in the body-free simulation at $Re=5{,}000$. The frequency spectra are averaged over 257 points in the spanwise direction, and the probability density functions are averaged over $\Delta T U_{\infty}/D>150$. ``Phase average'' denotes the case in which phase-averaged $\widetilde{U}$ and $\widetilde{V}$ extracted at $x/D=0.6$ from pre-computed DNS are used as the inflow boundary condition. The case $\alpha=1.0$ corresponds to time-averaged $\overline{U}$ extracted at $x/D=0.6$ and time-averaged $\overline{V}$ extracted at $x/D=1.4$ from the same DNS. In the remaining cases, the imposed inflow crossflow is obtained by rescaling $\overline{V}$ by the factor $\alpha$.}
       \label{fig:spectra_re5k}
\end{figure}

A similar strategy is applied to the body-free simulation at $Re=5{,}000$. In this case, $\overline{U}$ is extracted at $x/D=0.6$, while the reference $\overline{V}$ profile is extracted at $x/D=1.4$, where it exhibits peaks at $|y/D|=0.52$ with a maximum value of 0.34. Figure~\ref{fig:spectra_re5k} presents the frequency spectra and probability density functions obtained from the body-free simulation using the same $\overline{U}$ profile but different values of the imposed crossflow amplitude $\alpha\overline{V}$. The left panel shows that the frequency spectrum generated using the phase-averaged inflow is in good agreement with the DNS: both exhibit the same vortex-shedding frequency, a pronounced third harmonic, and a similar $-5/3$ scaling at high frequency. When only the time-averaged $\overline{U}$ is used, that is, for $\alpha=0$, the body-free simulation predicts a higher vortex-shedding frequency, although the third harmonic and the $-5/3$ scaling are still present. As $\alpha$ increases, corresponding to a stronger imposed crossflow, the vortex-shedding frequency shifts further upward. At $\alpha=2.5$, the flow exhibits features more typical of lower-Reynolds-number flow past a cylinder, as indicated by the appearance of additional harmonics in the spectrum, and at $\alpha=3.0$ no distinct vortex-shedding frequency can be identified. The pronounced change in wake behavior at $\alpha=2.5$ is further supported by the probability density function of the spanwise velocity gradient $\partial v/\partial z$, shown in the right panel of Figure~\ref{fig:spectra_re5k}, which develops a strongly non-Gaussian tail.

\begin{figure}
  \begin{subfigure}[t]{0.425\textwidth}
    \includegraphics[width=\textwidth,trim=10 20 200 10,clip]{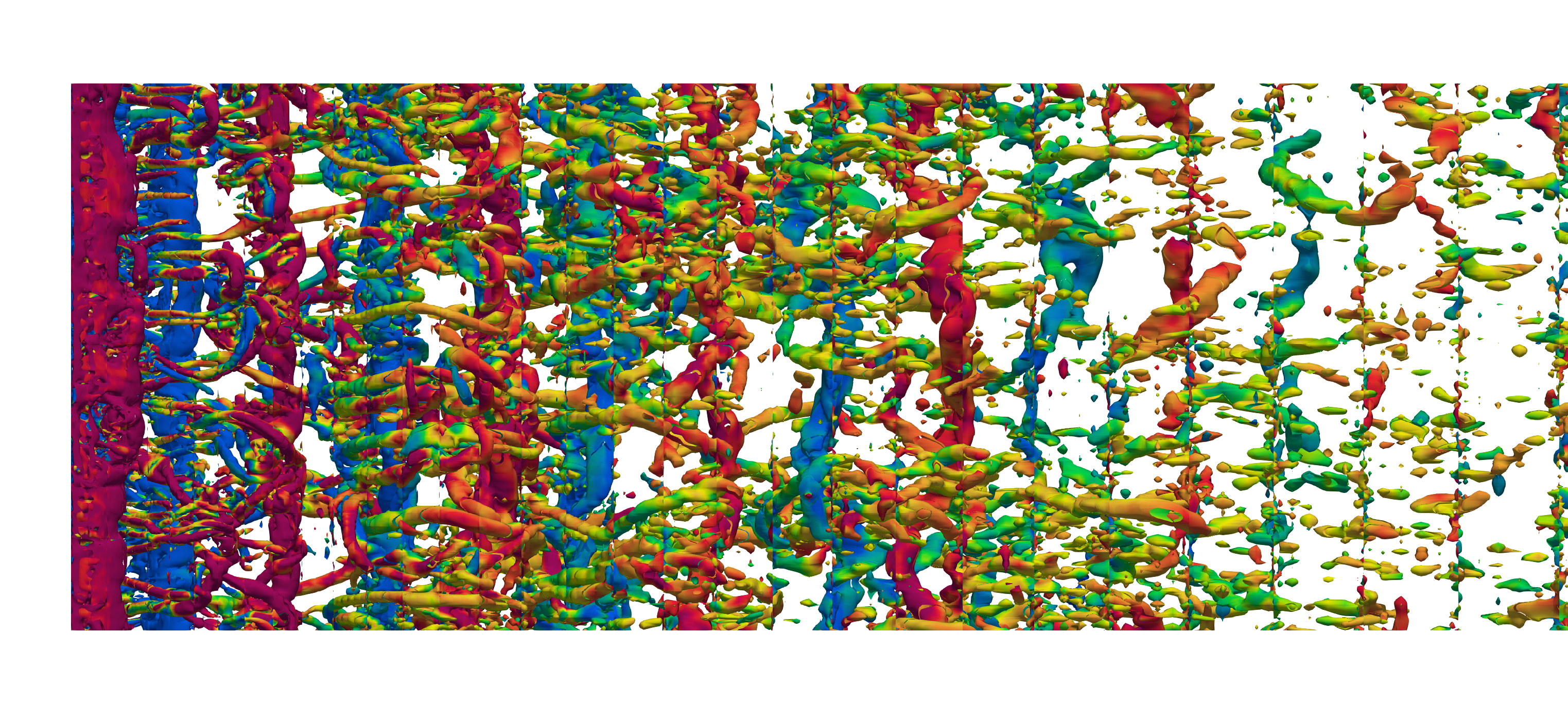}
        \label{fig:vort_phase_ave}
      \vspace*{-5mm}
    \subcaption*{Phase Averaged}
  \end{subfigure}
  \centering
    \hspace{5mm}
  \begin{subfigure}[t]{0.425\textwidth}
    \includegraphics[width=\textwidth,trim=10 20 200 10,clip]{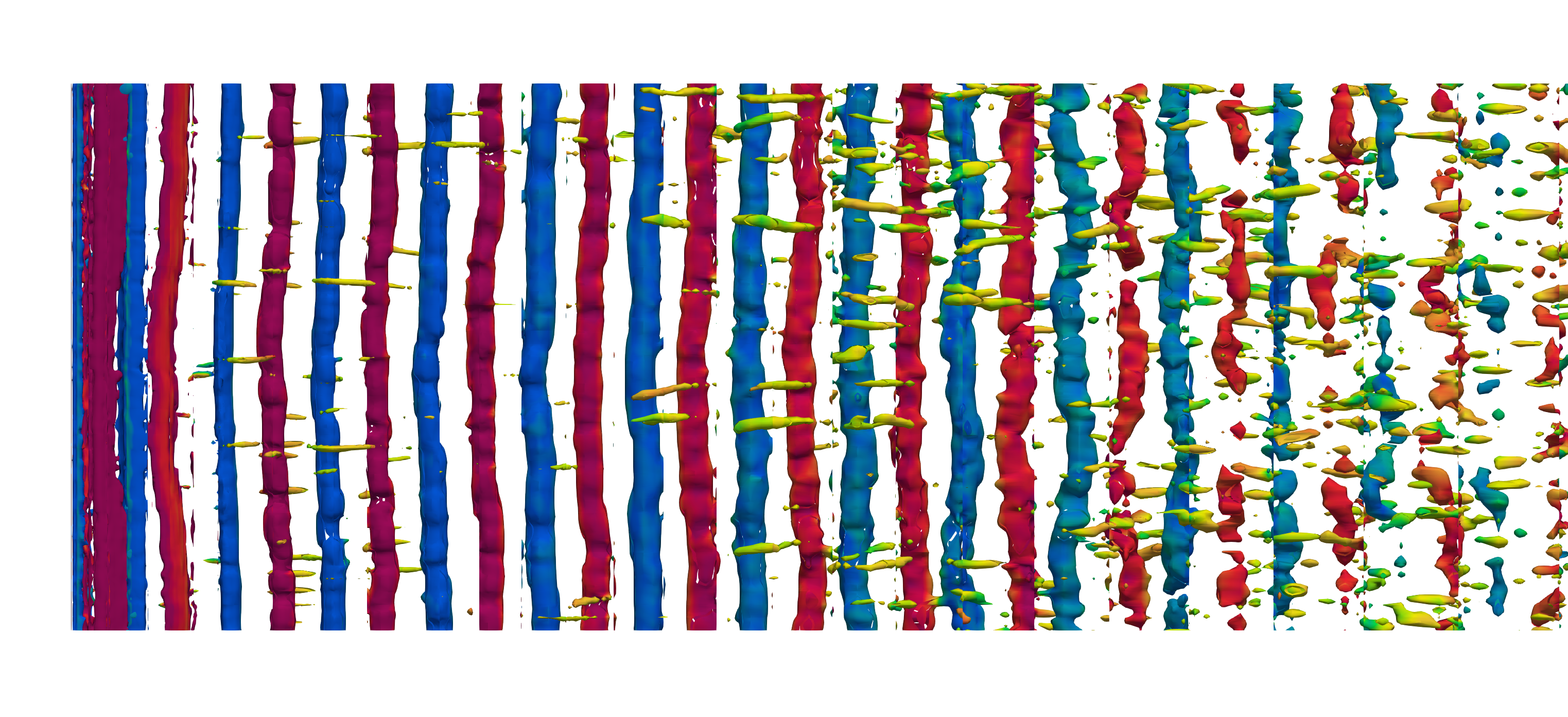}
    \label{fig:alpha2-5}
      \vspace*{-5mm}
     \subcaption*{ $\alpha=2.5$}
      \end{subfigure}   
   \caption{Wake structures from the body-free simulation at $Re=5,\,000$ for two representative inflow conditions. The vortices are visualized using iso-surfaces of $Q=0.5$, colored by the spanwise vorticity component $\omega_z$ in the range $[-2,2]$. The parameter $\alpha$ has the same meaning as in Figure~\ref{fig:spectra_re5k}.}
       \label{fig:re5k_vort3d}
\end{figure}

The significance of $\alpha\overline{V}$ in the body-free simulation can also be visualized through the wake structures shown in Figure~\ref{fig:re5k_vort3d}. When the phase-averaged data are used, the vortices generated by the body-free simulation exhibit clear features of three-dimensional turbulence. In contrast, for $\alpha=2.5$, the vortices display characteristics more typical of lower-Reynolds-number flow past a cylinder.
~

Before closing this section, we note the connection between the present study and recent investigations of bluff-body wake control using surface-mounted slot jets \cite{PRL_Dong}. Related approaches involving synthetic jets, plasma actuators, pulsed blowing, and small rotating cylinders have likewise been shown to suppress vortex shedding, reduce drag, and alter wake topology behind cylinders and other bluff bodies \cite{Mittal2005_AIAA,choi2008review,FanPNAS2020}. Taken together, these studies emphasize that modifying the near-wake momentum---whether through external actuation or through prescribed inflow conditions, as in the present body-free simulation framework---can fundamentally change the instability characteristics and transition pathways of bluff-body wakes.

More broadly, the present results suggest that the essential role of the upstream body is to generate an instability-bearing near-wake profile, after which the downstream dynamics can evolve largely autonomously. In that sense, the body-free simulation provides not only a computationally cheaper surrogate for full DNS, but also a useful tool for isolating which aspects of the inflow are dynamically important for sustaining wake turbulence. \rev{For the present $Re=11{,}000$ case, the body-free simulation is estimated to provide a wall-clock speedup of about $40\times$, primarily because the absence of boundary-layer resolution permits a larger time step and fewer elements; in addition, it reduces the hardware requirement from two A6000 GPUs to one RTX3090. This perspective may be particularly valuable for reduced-order modeling, data assimilation, inverse design, and flow-control applications in which the near-wake state is easier to measure or prescribe than the full body-resolved flow.}

\section{Conclusion}
\label{sec:conclusion}
\rev{We have introduced a body-free simulation framework that reproduces three-dimensional turbulent wakes without explicitly resolving the upstream body. By prescribing a one-dimensional inflow velocity profile obtained from DNS or experiments, the simulation reconstructs the downstream wake and captures coherent vortices, Reynolds stresses, and spectral features in good agreement with full simulations and measurements. Our results support the conclusion that the wake dynamics are governed primarily by the instability of a properly selected near-wake profile rather than by the body geometry itself. Including both streamwise and crossflow velocity components is important for recovering the correct Strouhal number and preserving three-dimensionality. At the same time, from an experimental standpoint, the mean streamwise profile is far more practical to obtain or prescribe than a spatially varying crossflow component or a phase-averaged inflow synchronized with the shedding cycle. Systematic variations of the imposed inflow reveal a clear transition between three- and two-dimensional wakes governed by the crossflow amplitude. More broadly, this body-free paradigm provides a physically interpretable and computationally efficient route for reconstructing and controlling turbulent wakes directly from sparse experimental measurements.}

\rev{At the same time, the present study has a well-defined scope. The framework is demonstrated here for the wake of a stationary circular cylinder, and its success depends on prescribing an inflow profile extracted from the absolutely unstable near-wake region. If the inflow is specified too far downstream, where the mean profile has already lost the instability needed to sustain the wake, the body-free simulation no longer reproduces the correct dynamics. Moreover, because the body is absent, the approach is not intended to recover near-body quantities such as surface forces or boundary-layer details. Extending the method to other bluff-body geometries, moving or deforming bodies, and more complex inflow environments remains an important direction for future work.}




\section{Code and Data Availability}
The GitHub repository associated with this work is available at

\url{https://github.com/vanchisen/body-free-cylinder-wake}.

It contains the manuscript source, analysis scripts, plotting utilities, and processed datasets used to generate the main figures and tables. Large raw simulation and experimental datasets are not included in the repository and can be made available upon reasonable request.

\appendix
\section{\rev{Divergence-free transient forcing used during start-up}}

\rev{To accelerate the onset of unstable three-dimensional motions from the initially quiescent state, the body-free simulations employ a weak transient volumetric forcing during the interval $0 \le tU_{\infty}/D \le 20$. The forcing is chosen to be long-wavelength and divergence-free:}
\begin{equation}
\rev{\mathbf{f}(x,y,z,t)=\hat{\mathbf{f}}(x,y,z)\sin(\omega t),}
\end{equation}
\rev{where}
\begin{equation}
\rev{\hat f_x=A\cos(k_x x)\sin(k_y y)\cos(k_z z),}
\end{equation}
\begin{equation}
\rev{\hat f_y=A\sin(k_x x)\cos(k_y y)\cos(k_z z),}
\end{equation}
\rev{and}
\begin{equation}
\rev{\hat f_z=A\frac{k_x+k_y}{k_z}\sin(k_x x)\sin(k_y y)\sin(k_z z).}
\end{equation}
\rev{Here,}
\begin{equation}
\rev{k_x=\frac{8\pi}{L_x},\qquad
k_y=\frac{8\pi}{L_y},\qquad
k_z=\frac{8\pi}{L_z},}
\end{equation}
\rev{$L_x$, $L_y$, and $L_z$ denote the streamwise, crossflow, and spanwise domain lengths, respectively, $A=0.1$, and $\omega=\pi$. By construction,}
\begin{equation}
\rev{\nabla\cdot\mathbf{f}=0.}
\end{equation}
\rev{This forcing is used only to seed unstable three-dimensional motions during start-up; all statistics reported in the main text are obtained from the subsequent unforced evolution.}

\section{Declaration of Interests}
The authors report no conflict of interest.

\section{Acknowledgments}
This research was supported by the Defense Advanced Research Projects Agency (DARPA) through the Automated Prediction Aided by Quantized Simulators (APAQuS) program under Grant No.~HR00112490526.
\bibliographystyle{elsarticle-num}
\bibliography{jcp-instructions}

\end{document}